\documentclass[useAMS,usenatbib]{mn2e}
\usepackage{graphicx}
\usepackage{times}
\newcommand{\mdev}{$m_{dev}$}
\newcommand{\mdevr}{$m_{dev,r}$}
\newcommand{\mpet}{$m_{p}$}
\newcommand{\mpetr}{$m_{p,r}$}
\newcommand{\photo}{{\it Photo}}
\newcommand{\lre}{$\log r_{e}$}
\newcommand{\re}{$r_{e}$}
\newcommand{\lsn}{$\log n$}
\newcommand{\mie}{$<\! \mu\! >_{e}$}
\newcommand{\slre}{$\sigma_{_{\log r_e}}$}
\newcommand{\smie}{$\sigma_{_{<  \!  \mu  \!  >_e}}$}
\newcommand{\sln}{$\sigma_{_{\log n}}$}

\newcommand{\slnlre}{$COV_{_{\log n,\log r_e}}$}
\newcommand{\slremie}{$COV_{_{\log r_e,<\! \mu\! >_{e}}}$}
\newcommand{\smieln}{$COV_{_{<\! \mu\! >_{e},\log n}}$}
\newcommand{\ls}{$\log \sigma_0$}

\newcommand{\mr}{$^{0.07}M_{r}$}
\newcommand{\mx}{$^{0.07}M_{X}$}
\newcommand{\mrdev}{$^{0.07}M_{dev,r}$}

\newcommand{\mt}{$m_T$}

\title[SPIDER I  -- Sample and  data analysis]{Spheroid's Panchromatic Investigation   in  Different   Environmental  Regions   (SPIDER)  -- I.  Sample and  galaxy parameters  in  the \MakeLowercase{griz}${\rm YJHK}$ wavebands.}
\author[F. La Barbera et al.]{F. La Barbera$^{1}$\thanks{E-mail: labarber@na.astro.it (FLB); rrdecarvalho2008@gmail.com(RRdC)}, R.R. de Carvalho$^{2}$, I.G. de la Rosa$^{3}$, P.A.A. Lopes$^{4}$, \newauthor
J.L.Kohl-Moreira$^{5}$ and H.V. Capelato$^{2}$ \\
$^1$INAF -- Osservatorio Astronomico di Capodimonte, Napoli, Italy \\
$^2$Instituto Nacional de Pesquisas Espaciais/MCT, S. J. dos Campos, Brazil\\
$^3$Instituto de Astrofisica de Canarias, Tenerife, Spain\\
$^4$Observat\'orio do Valongo/UFRJ, Rio de Janeiro, Brazil\\
$^5$Observat\'orio Nacional/MCT, Rio de Janeiro, Brazil}

\begin{document}
\date{Submitted on 2009 December 22}

\pagerange{\pageref{firstpage}--\pageref{lastpage}} \pubyear{2010}

\maketitle

\label{firstpage}
\begin{abstract}
This  is  the  first  paper   of  a  series  presenting  a  Spheroid's
Panchromatic   Investigation   in   Different  Environmental   Regions
(SPIDER).   The   sample  of   spheroids  consists  of   5,080  bright
($M_r<-20$) Early-Type galaxies (ETGs),  in the redshift range of 0.05
to  0.095,  with  optical  (griz)  photometry  and  spectroscopy  from
SDSS-DR6  and Near-Infrared (YJHK)  photometry from  UKIDSS-LAS (DR4).
We describe how homogeneous  photometric parameters (galaxy colors and
structural  parameters) are  derived using  $grizYJHK$  wavebands.  We
find  no systematic  steepening of  the CM  relation when  probing the
baseline from g-r to g-K, implying that internal color gradients drive
most of  the mass-metallicity relation  in ETGs. As far  as structural
parameters are  concerned we  find that the  mean effective  radius of
ETGs  smoothly  decreases,  by  ~30\%,  from g  through  K,  while  no
significant  dependence on waveband  is detected  for the  axis ratio,
Sersic index, and a$_{\rm  4}$ parameters.  Also, velocity dispersions
are re-measured for all the ETGs using STARLIGHT and compared to those
obtained by SDSS.   { The velocity dispersions are  re-derived using a
  combination of simple stellar  population models as templates, hence
  accounting   for  the   kinematics  of   different   galaxy  stellar
  components.}   We   compare  our  (2DPHOT)   measurements  of  total
magnitude, effective  radius, and  mean surface brightness  with those
obtained  as   part  of   the  SDSS  pipeline   (Photo).   Significant
differences are found and reported, including comparisons with a third
and  independent   part.   A  full  characterization   of  the  sample
completeness in all wavebands is presented, establishing the limits of
application of  the characteristic  parameters presented here  for the
analysis of the global scaling relations of ETGs.
\end{abstract}

\section{Introduction}
\label{sec:INTR}

We have witnessed tremendous advances in our ability to measure galaxy
properties with unprecedented accuracy  over the past two decades. The
greatest leap  forward was  the advent  of the CCD,  with an  order of
magnitude  increase  in sensitivity  over  photographic  film and  the
ability to easily make  quantitative measurements. Today, surveys like
the Sloan Digital Sky Survey (SDSS) and UKIRT Infrared Deep Sky Survey
(UKIDSS)  provide  access  to  high  quality  data  covering  a  large
wavelength  range,  probing  different  astrophysical aspects  of  the
galaxies.  These  advances have been immediately  applied to examining
the global properties of galaxies.

Understanding the formation and evolution of galaxies requires probing
them over  a long time  (redshift) baseline to establish  the physical
processes responsible  for their  current observed properties.   It is
far easier to measure nearby  ETGs as opposed to their counterparts at
high redshift, and  this must be borne in  mind when comparing samples
at opposite  distance extremes.  Almost  as soon as high  quality data
became available  for nearby  samples of ETGs  it was  recognized that
they  occupy  a  2-dimensional  space,  the  fundamental  plane  (FP),
represented by the quantities radius, velocity dispersion, and surface
brightness  R - $\sigma$  - $\mu$.   Brosche (1973)  was the  first to
examine  ETGs  using  multi-variate statistical  techniques,  applying
Principal  Component  Analysis  (PCA)  to  the  relatively  poor  data
available.   Although his results  were not  fully appreciated  at the
time, it drew  the attention of other researchers  who further studied
the implications  of the FP  (Djorgovski 1987; Dressler et  al. 1987).
Brosche's  fundamental contribution  was  to show  that  we should  be
looking  for sets  of data  with  the smallest  number of  significant
principal  components  when starting  from  a  large  number of  input
parameters. In this  way we reduce a high-dimensional  dataset to only
those  quantities that are  likely to  have physical  meaning. Several
studies followed  Brosche's: \citet{Bujarrabal:81},  Efstathiou \&
Fall~(1984), Whitmore~(1984), and Okamura et al.~(1984).

The fundamental plane is a  bivariate scaling law between R, $\sigma$,
and $\rm  I$, where  $\mu$ = -2.5  log $\rm  I$, expressed as  R $\sim
\sigma^{\rm  A} \rm  I^{\rm  B}$.   In order  to  obtain accurate  and
meaningful   coefficients  that   can  be   compared   to  theoretical
expectations (such as those from the Virial Theorem, which implies A =
2 and B=  -1) we need to  not only have a homogeneous  sample of ETGs,
but also  to understand the  selection effects in defining  the sample
and to  properly measure the photometric  and spectroscopic quantities
involved. Several  contributions in  the past have  met most  of these
requirements~\citep{GrC97, JFK96}.   However, the lack  of homogeneous
data  covering a  large  wavelength baseline  while  also probing  the
entire range  of environments  (local galaxy density)  impeded further
progress.  Bernardi et  al. (2003a,b),  Bernardi et  al.   (2006), and
Bernardi et al. ( 2007) were the first to fill this gap and set strong
constraints  on the  FP  coefficients and  their implications,  though
limited   to  the   optical  regime   (see  also   Hyde   \&  Bernardi
2009). Another  important and often overlooked aspect  of such studies
is  the  impact  of   different  techniques  and  implementations  for
measuring  R -  $\sigma$ -  $\mu$ and  their respective  errors, which
ultimately  will be  propagated and  compared to  the  distribution of
residuals around the FP~\citep{PrS96, Gargiulo:09}.

This  is  the  first  paper  of a  series  presenting  the  Spheroid's
Panchromatic Investigation in Different Environmental Regions (SPIDER)
survey. SPIDER utilizes optical  and Near-Infrared (NIR) photometry in
the $grizYJHK$ wavebands as  well as spectroscopic data.  Spectroscopy
and optical photometry are taken  from SDSS DR6, while the $YJHK$ data
are obtained from  the UKIDSS-LAS DR4. In the  present work (Paper I),
we describe  how the sample of  ETGs is selected,  how the photometric
and spectroscopic  parameters are derived for each  galaxy, and derive
an  accurate  estimate of  the  completeness  of  the sample  in  each
band. The  $grizYJHK$ galaxy  images have been  homogeneously analyzed
using 2DPHOT, an automatic software designed to obtain both integrated
and     surface    photometry     of     galaxies    in     wide-field
images~\citep{LBdC08}. We present a  detailed comparison of the 2DPHOT
output  quantities  (magnitudes and  structural  parameters) to  those
provided  by  the  SDSS  Photo  pipeline~\citep{EDR}.   We  have  also
re-computed central  velocity dispersions from the  SDSS spectra using
the software STARLIGHT~\citep{CID05}, and compared these new estimates
to  those from  SDSS. {  Velocity dispersions  are re-derived  using a
  combination of  simple stellar population models  as templates. This
  procedure  minimizes  the  well  known  template  mismatch  problem,
  accounting   for  the  different   kinematics  of   various  stellar
  components  in  a galaxy.   All  the  photometric and  spectroscopic
  measurements  presented here  are  made available  through an  ascii
  table                                                              at
  http://www.lac.inpe.br/bravo/arquivos/SPIDER\_data\_paperI.ascii.\footnote{The
    file is  mirrored at http://www.na.astro.it/~labarber/SPIDER/.}  }
{ The  complete SPIDER  data-set is also  made available,  on request,
  through  a  database  structure  which  allows the  user  to  easily
  retrieve all information by issuing SQL queries}.

In Sec.~\ref{sec:DATA}, we describe how the galaxy sample is selected.
Sec.~\ref{sec:PHOT} describes how  $grizYJHK$-band images are analyzed
to derive  integrated photometry and the  structural parameters, with
corresponding uncertainties. Secs.~\ref{sec:CM} and~\ref{sec:dist_par}
compare the overall integrated  and structural properties of ETGs from
$g$  through $K$, deriving  color--magnitude relations  and presenting
the   distribution  of   structural  parameters   in   all  wavebands.
Sec.~\ref{sec:conf_SDSS_2DPHOT}  compares  the  structural  parameters
derived from  2DPHOT with those from  SDSS. In Sec.~\ref{sec:spectros},
we describe the measurement of central velocity dispersions, comparing
them to those from SDSS. The  completeness of the sample is studied in
Sec.~\ref{sec:compl}. A summary is provided in Sec.~\ref{sec:summary}.

Throughout the paper, we adopt a cosmology with $\rm H_0 \! = \! 75 \,
km  \,  s^{-1}  \, Mpc^{-1}$,  $\Omega_{\rm  m}  \!  = \!   0.3$,  and
$\Omega_{\Lambda} \! = \! 0.7$.

\section{Sample selection}
\label{sec:DATA}

\subsection{Sample Definition}
\label{sec:samp_def}

The sample  of ETGs is  selected from SDSS-DR6, following  a procedure
described  in~\citet{LBM08} and~\citet{LdC09},  selecting  galaxies in
the redshift  range of 0.05 to 0.095,  with $^{0.1}M_{r}{<}-20$, where
$^{0.1}M_{r}$ is  the k-corrected SDSS Petrosian  magnitude in r-band.
The k-correction  is estimated using the  software $kcorrect$ (version
$4\_1\_4$;  ~\citealt{BL03a},  hereafter  BL03), through  a  restframe
r-band   filter  blue-shifted   by  a   factor  $(1+z_0)$   (see  also
Sec.~\ref{sec:int_phot}).        As       in      previous       works
(e.g.~\citealt{Hogg04}), we adopt $z_0=0.1$.  The lower redshift limit
of the sample is chosen to minimize the aperture bias~\citep{GOMEZ03},
while  the  upper  redshift  limit  guarantees (1)  a  high  level  of
completeness  (according  to~\citealt{SAR06})  and  (2) allows  us  to
define a volume-limited sample  of {\it bright} early-type systems. In
fact,  ETGs  follow  two  different  trends  in  the  size--luminosity
diagram~\citep{capaccioli1992,   graham&guzman2003}.   The  separation
between  the   two  families  of  {\it  bright}   and  {\it  ordinary}
ellipticals  occurs  at  an  absolute  B-band  magnitude  of  $  -19$,
corresponding to the magnitude limit  of $M_r \sim -20$ we adopt here.
At the upper  redshift limit of $z=0.095$, the  magnitude cut of $-20$
also corresponds  approximately to the magnitude limit  where the SDSS
spectroscopy  is complete  (i.e. a  Petrosian magnitude  of  $m_r \sim
17.8$).  Following  ~\citet{BER03a}, { we  define ETGs using  the SDSS
  spectroscopic parameter  $eClass$, that indicates  the spectral type
  of a galaxy on the basis  of a principal component analysis, and the
  SDSS photometric parameter  $fracDev_r$, which measures the fraction
  of galaxy  light that is better  fitted by a  de Vaucouleurs (rather
  than  an exponential)  law.  In  this contribution,  ETGs  are those
  systems with $eClass  \!  < \!  0$ and $fracDev_r \!   > \!  0.8$. }
We select only galaxies  with central velocity dispersion, $\sigma_0$,
available from SDSS-DR6, in the  range of $70$ and $420$ km\,s$^{-1}$,
and with  no spectroscopic warning on (i.e.   $zWarning$ attribute set
to zero).   These constrains  imply retrieving only  reliable velocity
dispersion measurements from  SDSS.  All the above criteria  lead to a
sample of $39,993$ ETGs.

\begin{figure}
\begin{center}
\includegraphics[width=80mm,height=80mm]{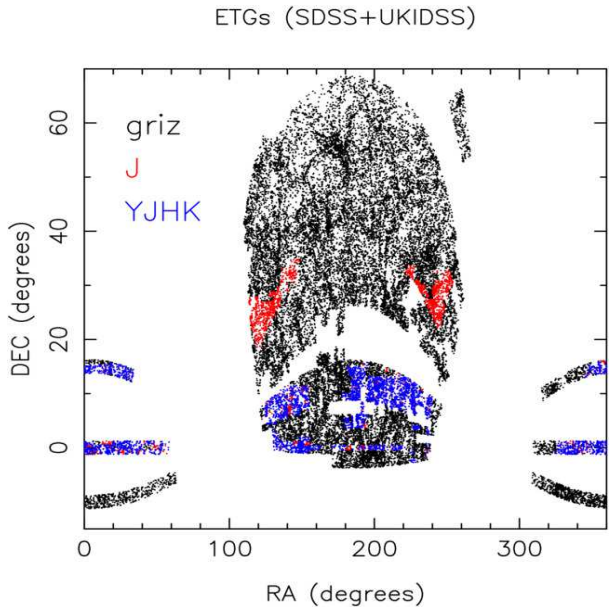}
\caption{Distribution in  $RA$ and $DEC$  of the SPIDER  sample. Black
  points  mark the  optical $griz$  data, while  red and  blue symbols
  denote the $J$ and $YJHK$ data, respectively.  .~\label{sky}}
\end{center}
\end{figure} 

The sample  of ETGs with  optical data is  then matched to  the fourth
data release (DR4) of UKIDSS--Large Area Survey (LAS). UKIDSS--LAS DR4
provides NIR  photometry in the  $YJHK$ bands over $\sim  1000$ square
degrees on the sky,  with significant overlap with SDSS~\citep{Law07}.
The $YHK$-band data have a  pixel scale of $0.4 \, ''/pixel$, matching
almost  exactly   the  resolution  of  the  SDSS   frames  ($0.396  \,
''/pixel$).  $J$  band observations are carried out  with a resolution
of $0.4  \, ''/pixel$, and then  interleaved to a  subpixel grid. This
procedure results into stacked frames with a better resolution of $0.2
\,  ''/pixel$. The  $YJHK$ stacked  images (multiframes)  have average
depths~\footnote{defined  by  the  detection  of  a  point  source  at
  $5\sigma$ within a 2$''$  aperture.}  of $20.2$, $19.6$, $18.8$, and
$18.2$~mags,  respectively. The matching  of SDSS  to UKIDSS  data was
done with  the {\it  CrossID} interface of  the WFCAM  Science Archive
website~\footnote{See    http://surveys.roe.ac.uk/wsa/index.html   for
  details.}.  For  each ETG  in the SDSS  sample, we searched  for the
nearest UKIDSS detection within a radius of $1$'', by considering only
UKIDSS frames with better quality flag ($ppErrBits <16$). The matching
result was very  insensitive to the value of  the searching radius. In
fact,  changing  it to  $0.5$''  leads  to  decrease the  sample  with
galaxies having $YJHK$ data available by only five objects. The number
of matched sources is maximum in $J$ band, with $7,604$ matches, while
amounts  to 5,698,  6,773, and  6,886 galaxies  in $Y$,  $H$,  and $K$
bands, respectively.  Considering ETGs simultaneously matched with two
UKIDSS bands,  $H \! +  \! K$ provides  the maximum number  of objects
($6,575$).  For any possible set of three bands, the number of matches
varies between  $5,228$ ($Y+J+K$) to  $5,323$ ($J+H+K$), which  is not
significantly larger than the number of $5,080$ ETGs having photometry
available in all the $YJHK$ bands.  For this reason, we have retrieved
NIR data for  only those galaxies with available  photometry in either
$J$  ($7,604$), or  $H \!   + \!   K$ ($6,575$),  or  $YJHK$ wavebands
($5,080$).   The   completeness  in   magnitude  of  each   sample  is
characterized in Sec.~\ref{sec:compl}.

In summary, the SPIDER  sample includes 39,993 galaxies with available
photometry  and  spectroscopy  from  SDSS-DR6.   Out  of  them,  5,080
galaxies have NIR photometry in DR4 of UKIDSS-LAS. The distribution of
galaxies  with optical  and  NIR data  on  the sky  is illustrated  in
Fig.~\ref{sky}.

\subsection{Contamination by Faint Spiral Structures}
\label{sec:cont_spiral}

\begin{figure*}
\begin{center}
\includegraphics[width=160mm,height=150mm]{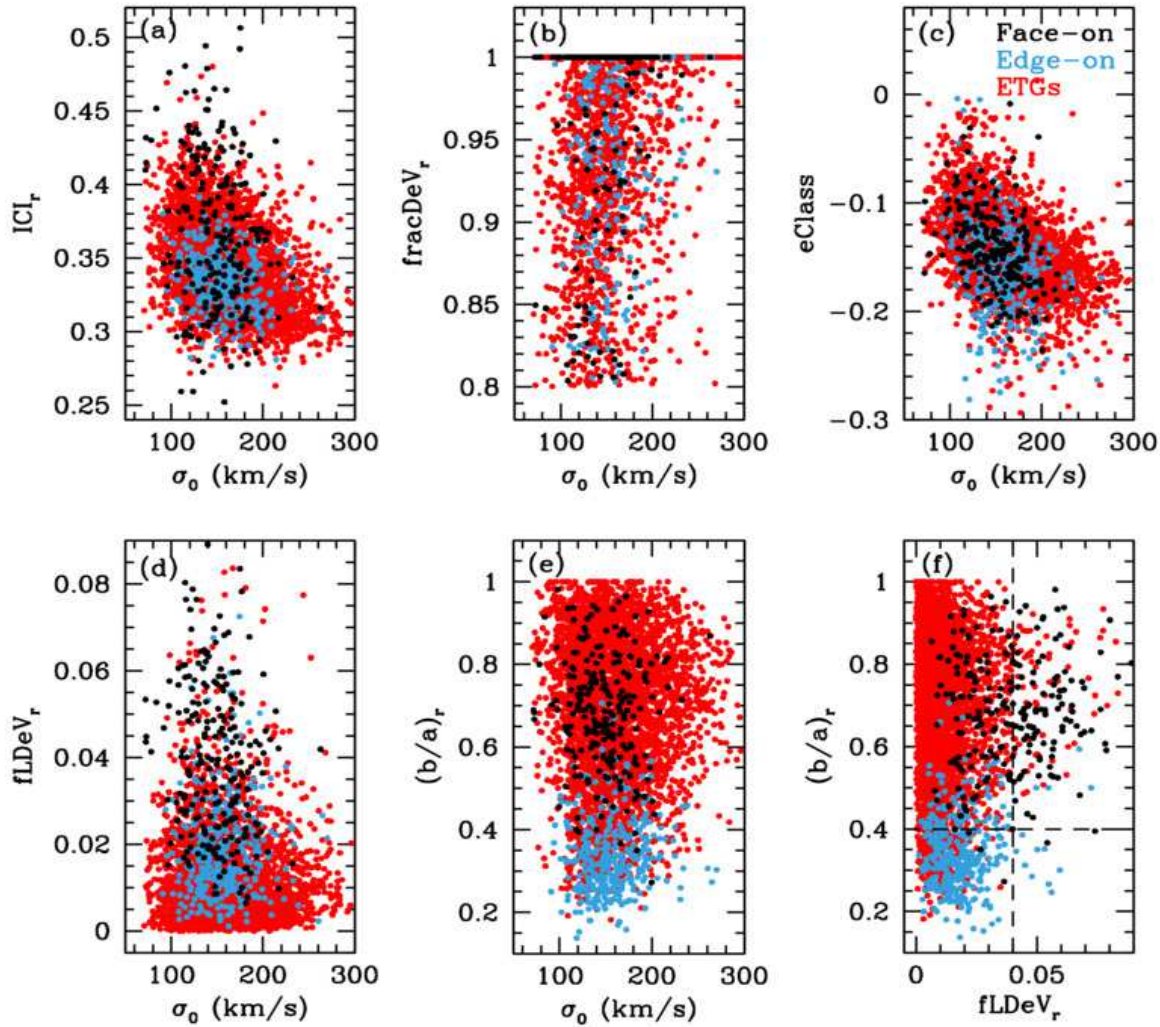}
\caption{Different morphological indicators  are plotted as a function
  of the central velocity dispersion  of the galaxy for a subsample of
  4,000 ETGs (panels a-e). Panel  (f) plots the two optimal indicators
  and the dashed lines indicate  the region where the contamination is
  $\sim$6\% ((b/a)$_r  > 0.4$ and  fLDeV$_r <$ 0.04. Face-on  LTGs are
  indicated in black,  edge-on LTGs in blue and  bonafide ETGs in red,
  as    shown   in    the   upper-right    corner   of    panel   (d).
  ~\label{bonafide1}}
\end{center}
\end{figure*}

{
Specification of  a given family  of systems means setting  a property
(or  properties) that  isolates systems  that presumably  went through
similar  evolutionary  processes.  However,  when  we  select as  ETGs
systems with SDSS parameters $eClass \!   < \!  0$ and $fracDev_r \! >
\!   0.8$ we  expect a  certain  amount of  contamination by  galaxies
exhibiting   faint  structures   resembling  spiral   arms   or  other
non-systemic morphologies.   Although several morphological indicators
have been  proposed from  the parameters of  the SDSS  pipeline (e.g.
Strateva et al. 2001),  the eyeball classification is still considered
one of  the most reliable  indicator despite its  evident subjectivity
(Weinmann et  al.  2009).  We  have visually inspected a  subsample of
4,000 randomly chosen galaxies  from our sample, classifying them into
three groups: {\it ETGs} (featureless spheroids); {\it face-on} LTGs -
late type galaxies  (a bulge surrounded by an  obvious disk); and {\it
  edge-on} LTGs (a bulge  with a prominent disk).  This classification
is then  used to evaluate  the ability of the  different morphological
indicators  to  distinguish  edge-on  and face-on  galaxies  from  the
bonafide ETGs.

Five  SDSS  morphological  indicators   are  considered  as  shown  in
Figs.~\ref{bonafide1}(a-e): (i) The r-band Inverse Concentration Index
(ICI$_{\rm r}$),  defined by the ratio  of the 50\%  to the 90\%-light
Petrosian  radii (see  Shimasaku  et al.   2001);  (ii) The  parameter
$fracDeV_r$, corresponding  to the fraction of the  total fitted model
accounted for the de Vaucouleurs component.  Notice that the total fit
is  not  a bulge$+$disk  decomposition  but  a  re-scaled sum  of  the
best-fitting  de Vaucouleurs and  exponential components  (Bernardi et
al. 2006); (iii) The $eClass$ indicator (see Sec.~\ref{sec:samp_def});
(iv)  The  fractional  likelihood   of  a  de  Vaucouleurs  model  fit
(fLDeV$_r$), defined as:
\begin{equation}
fLDeV_r = \frac{lnLDeV_r}{(lnLDeV_r+lnLexp_r+lnLstar)}
\label{eq:fLDeV}
\end{equation}
\noindent where  $LDeV$, $Lexp$ and  $Lstar$ are the  probabilities of
achieving the measured chi-squared for the de Vaucouleurs, exponential
and PSF fits, respectively; and (v) The projected axis ratio (b/a)$_r$
of  a  deVaucouleurs  fit  (deVAB$_r$  -  an  SDSS  attribute).   From
Fig.~\ref{bonafide1}  fLDeV$_r$  and  (b/a)$_r$  are  the  two  better
performance  indicators  discriminating ETGs  from  face-on LTGs  (see
Maller et al.   2009).  Based on the visual  inspection, we define two
cutoff values  (see panel  f in the  Figure), one for  each morphology
indicator,  defining   a  region  where  the   contamination  rate  is
$\sim$5\%.  We notice that  this contamination rate is $\sim$2.5 times
smaller than that  for the sample of $4,000$.  The selected values for
the   cutoffs  are  0.04   and  0.4   for  fLDeV$_r$   and  (b/a)$_r$,
respectively.  The same constraints  imposed to the entire ETG sample,
define a sub-sample of 32,650 bonafide ETGs.  We flag 7,343 objects as
lying in  the non-bonafide ETGs region  defined by (b/a)$_r  < 0.4$ or
fLDeV$_r >$ 0.04,  so that we can study the  impact of contaminants on
the global properties of bonafide ellipticals.  }

\section{Photometry}
\label{sec:PHOT}

\begin{figure}
\begin{center}
\includegraphics[width=85mm,height=85mm]{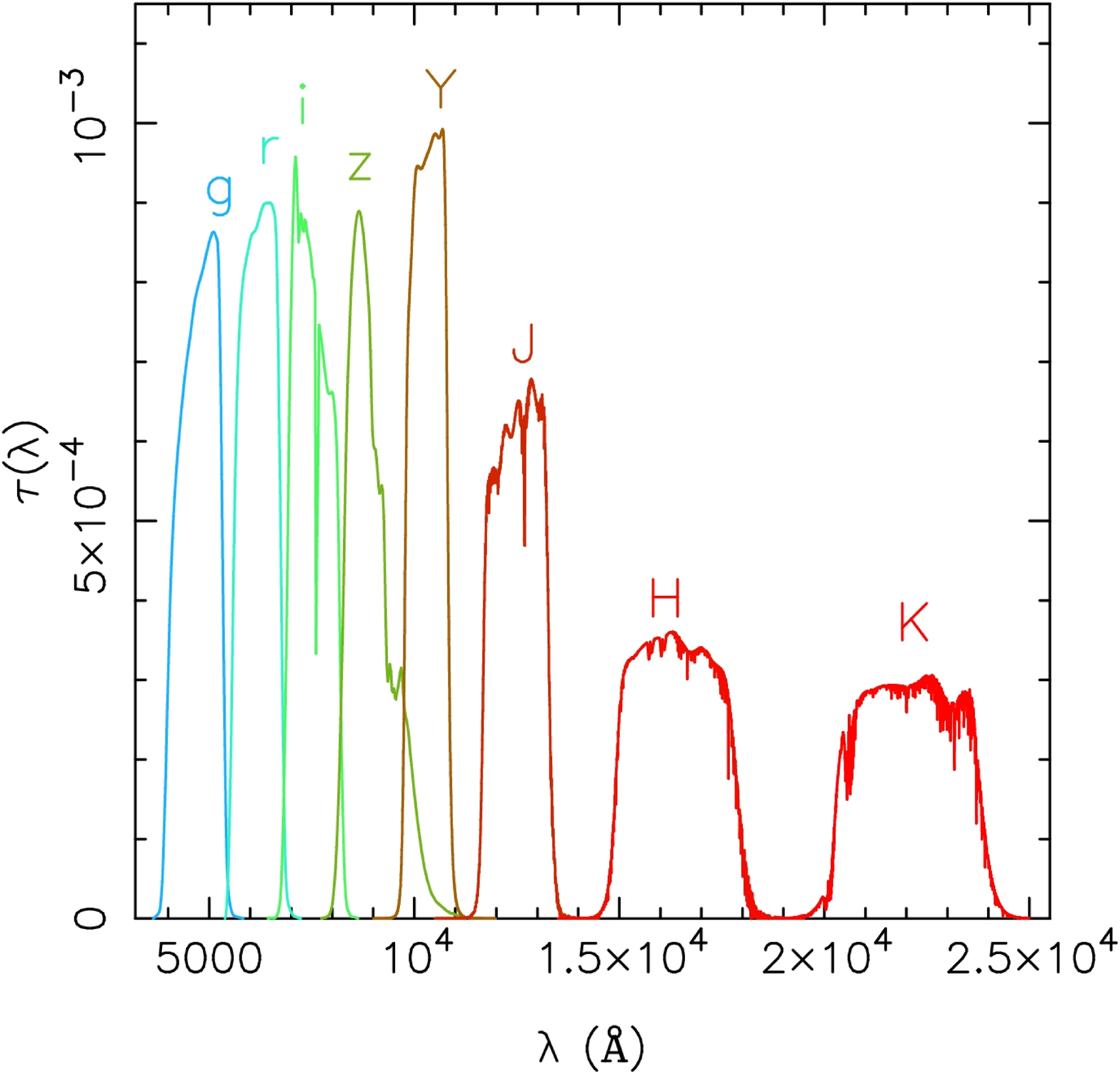}
\caption{Transmission curves, $\tau(\lambda)$,  for the $grizYJHK$ filters.
Each curve has been normalized to an area of one.
~\label{filters}}
\end{center}
\end{figure}

For each  ETG, we  retrieved the corresponding  best-calibrated frames
from  the SDSS  archive and  the  multiframes from  the WFCAM  Science
Archive.  In  the case  of SDSS, only  the griz images  were analyzed,
since the  signal-to-noise of  the u-band data  is too low  to measure
reliable  structural  parameters.  The resulting  photometric  system,
consisting    of    the   $grizYJHK$    wavebands,    is   shown    in
Fig.~\ref{filters}, where  we plot,  for each band,  the corresponding
overall transmission curve.
Regarding photometric  calibration, the zero--point of  each image was
retrieved from the corresponding SDSS  or UKIDSS archives. As a result
of  the different  conventions adopted  in the  two surveys,  the SDSS
photometry  is in  the $AB$  photometric  system~\footnote{We actually
  apply small offsets to the  $griz$ zero-points in order to produce a
  better   match   between   the    SDSS   and   $AB$   systems   (see
  www.sdss.org/dr6/algorithms/fluxcal.html\#sdss2ab).},  while  UKIDSS
data are calibrated into the Vega system~\citep{Law07}.

\subsection{Integrated properties}
\label{sec:int_phot}
We  have  measured both  aperture  and  total  galaxy magnitudes  with
different methods,  homogeneously for both  the optical and  NIR data.
Aperture  magnitudes  are  estimated with  S-Extractor~\citep{BeA:96}.
For each galaxy,  several apertures are measured, spanning  the $2$ to
$250$ pixels diameter range.  A set of adaptive aperture magnitudes is
also measured.  The adaptive  apertures have diameters of $d_k=k \cdot
r_{_{Kr,i}}$, where  $k=3, 4, 5 ,6  $ is a  multiplicative factor, and
$r_{_{Kr,i}}$ is  the Kron radius  ~\citep{Kron:80} in the  i-band, as
estimated with  S-Extractor. The  $r_{_{Kr,i}}$ is measured  in i-band
since this  is approximately in  the middle of the  SPIDER photometric
system (Fig.~\ref{filters}).  For $k=3$,  the median value of $d_k$ is
$\sim 5.8''$,  which is more than  five times larger  than the typical
FWHM in  all wavebands  { (see Sec.~\ref{sec:struc_par})}.   More than
$95\%$ of all the ETGs have a value of $d_3$ larger than $\sim 4.8''$,
and a ratio  of $d_3$ to the seeing FWHM value  larger than $3$.  This
makes the adaptive aperture  magnitudes essentially independent of the
seeing variation from  $g$ through $K$ (see Sec.~\ref{sec:struc_par}).
Different  types of  total magnitudes  are adopted.   For  the optical
wavebands, we  retrieve both petrosian and model  magnitudes, \mpet \,
and  \mdev, from  the  SDSS archive  (see  ~\citealt{EDR}).  For  each
waveband, the Kron magnitude, $m_{Kr}$, is also measured independently
for all galaxies, within an aperture of three times the Kron radius in
that  band.   The  Kron   magnitudes  are  obtained  from  $MAG\_AUTO$
parameter as  estimated in S-Extractor.  Finally, each  galaxy has the
estimate   of   total   magnitude,   \mt,   from   the   corresponding
two-dimensional fitting model (see Sec.~\ref{sec:struc_par}).

To obtain  homogeneous measurements  from $g$ through  $K$, magnitudes
are k-corrected and dereddened for galactic extinction by re-computing
both corrections with the same procedure in all wavebands, rather than
retrieving   them,  when   available,   from  the   SDSS  and   UKIDSS
archives. For each galaxy, the  amount of extinction is estimated from
the reddening maps of Schlegel, Finkbeiner, and Davis (1998), applying
the correction  of Bonifacio, Monai  \& Beers (2000) that  reduces the
color  excess   value,  $E(B-V)$,   in  regions  of   high  extinction
($E(B-V)>0.1$). This correction is  not included in the SDSS database,
and only a  very small fraction of ETGs ($<1\%$) is  found in the high
extinction  regions.   We computed  k-corrections  using the  software
$kcorrect$ (BL03), through restframe filters obtained by blue-shifting
the throughput curves in Fig.~\ref{filters} by a factor $(1+z_0)$. For
$z_0=0$,  one  recovers  the  usual  k-correction.   For  galaxies  at
redshift $z=z_0$,  the k-correction is  equal to $-2.5  \log (1+z_0)$,
independent  of the  filter  and  the galaxy  spectral  type. We  have
adopted $z_0=0.0725$ which  is the median redshift of  the ETG sample.
According   to  BL03,   this  choice   allows  uncertainties   on  the
k-corrections to  be minimized~\footnote{The value  of $z_0=0.0725$ is
  smaller  than  that  of   $z_0=0.1$  adopted  for  sample  selection
  (Sec.~\ref{sec:DATA}).  The  value of $z_0=0.1$  makes the selection
  more similar to that performed from previous SDSS studies, while the
  choice of  $z_0=0.0725$ minimizes the errors  in k-corrections.}. We
have   tested    how   the   waveband   coverage    can   affect   the
k-corrections. For the sample  of $5,080$ galaxies with available data
in all $grizYJHK$ bands, we have estimated the k-corrections in $griz$
for two cases, where we used (i) all the eight wavebands and (ii) only
the  SDSS bands.   k-corrections turned  out  to be  very stable  with
respect to the adopted waveband's  set, with the standard deviation of
k-correction  differences being  smaller  than 0.01$mag$  for all  the
$griz$ wavebands.

\subsection{Structural parameters}
\label{sec:struc_par}
The  $grizYJHK$  images   were  processed  with  2DPHOT~\citep{LBdC08}
(hereafter  LdC08),  an automated  software  environment that  allows
several  tasks,  such   as  catalog  extraction  (using  S-Extractor),
star/galaxy separation,  and surface  photometry to be  performed. The
images were  processed using two  Beowulf systems. The  optical images
were  processed  at  the  INPE-LAC cluster  facility,  running  2DPHOT
simultaneously  on  40 CPUs.   A  number  of $31,112$  best-calibrated
frames were processed in each of the $griz$ wavebands, requiring $\sim
2$ days  per band.   The UKIDSS frames  were processed at  the Beowulf
system available  at INAF-OAC.  A  total of $12,963$  multiframes were
processed  by   running  2DPHOT  on  32   CPUs,  simultaneously.   The
processing took half a day for each band.

A complete  description of the 2DPHOT  package can be  found in LdC08;
here  we only  outline the  basic  procedure followed  to measure  the
relevant  galaxy parameters.   Both the  optical and  NIR  images were
processed  with  the same  2DPHOT  setup  to  guarantee a  homogeneous
derivation of  structural parameters from  $g$ through $K$.   For each
frame,  the  so-called  {\it  sure  stars}  are  identified  from  the
distribution   of  all   the  detected   sources  in   the   FWHM  vs.
signal-to-noise, $S/N$, diagram. This  procedure allows an estimate of
the average  seeing FWHM of  the image to  be obtained (see  sec.~3 of
LdC08). For each ETG, a local  PSF model is constructed by fitting the
four closest stars to that  galaxy with a sum of three two-dimensional
Moffat functions.  Deviations  of the PSF from the  circular shape are
modeled  by describing  the  isophotes of  each  Moffat function  with
Fourier-expanded   ellipses.    Galaxy   images   were   fitted   with
PSF-convolved Sersic  models having elliptical isophotes  plus a local
background  value.  For each  galaxy, the  fit provides  the following
relevant parameters:  the effective (half-light)  radius, $r_{\rm e}$,
the mean  surface brightness within  that radius, $<  \!\mu\!>_{\rm e,
}$, the Sersic index (shape  parameter) $\rm n$, the axis ratio $b/a$,
and the position angle of  the major axis, $PA$.  The total (apparent)
magnitude, $m_T$,  of the model  is given by the  definition $m_T=-2.5
\log (2 \pi)-5 \log (r_e) +<  \! \mu \! >_e$.  Mean surface brightness
values  are  k-corrected  and  corrected for  galactic  extinction  as
described  in Sec.~\ref{sec:int_phot}. Moreover,  cosmological dimming
is removed by  subtracting the term $10 \log (1+z)$,  where $z$ is the
SDSS spectroscopic redshift.

The characterization of the galaxy isophotal shape is done through the
two-dimensional  fitting of  each ETG  in the  $gri$  wavebands, where
Sersic models having  isophotes described by Fourier-expanded ellipses
are adopted. Only the fourth order $cos$ term of the expansion, $a_4$,
is  considered  (boxiness  -  $a_4<0$  and diskyness  -  $a_4>0$,  see
e.g.~\citealt{Bender:87}, hereafter BM87).  Only the $gri$ band images
are  analyzed,  since  $a_4$   is  usually  measured  in  the  optical
wavebands.  Since  the  models  are PSF-convolved,  the  method  above
provides  a global deconvolved  estimate of  $a_4$.  This  estimate is
somewhat different from  the definition of BM87, where  the peak value
of $a_4$ is  derived in a given radial range,  with the minimum radius
being set  to four  times the  seeing FWHM and  the maximum  radius to
twice the effective  radius. Since many galaxies in  the SPIDER sample
have effective radii  comparable to a few times  the seeing FWHM value
(see below), the BM87 procedure is not applicable.

Fig.~\ref{seeing} shows  the distribution  of the average  seeing FWHM
value for  all the retrieved images  from $g$ through  $K$. The seeing
FWHM was  estimated from the {\it  sure star} locus  (see above).  For
each band, we estimate the  median of the distribution of FWHM values,
and   the   corresponding    width   values,   using   the   bi-weight
statistics~\citep{Beers:90}. The median  and width values are reported
in Fig.~\ref{seeing} for each band. As expected, the median FWHM value
tends to  smoothly decrease  from the blue  to NIR  wavebands, varying
from  $1.24''$  in  g-band  to  $0.82''$ in  K-band.   This  variation
corresponds to a  relative change of $\sim 34\%$  (with respect to $g$
band). Notice  also that  in $YJHK$ ($griz$)  bands almost  all frames
have seeing FWHM values better  than $1.5''$ ($1.8''$), with $90\%$ of
the values  being smaller than $1.2''$ ($1.5''$).   This decreasing of
the  seeing FWHM  from  $g$  through $K$  matches  almost exactly  the
relative change of effective radii  from optical to NIR wavebands (see
below), making the ratio of FWHM to $r_e$ almost constant from g to K.
\begin{figure}
\begin{center}
\includegraphics[height=85mm]{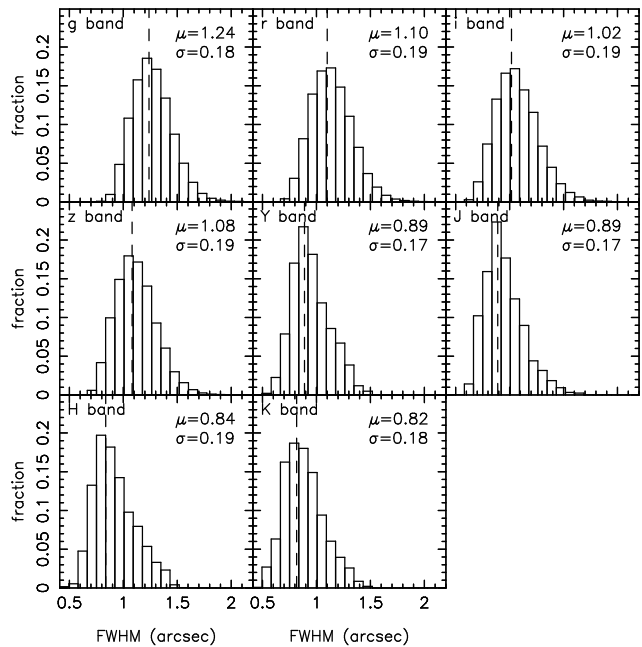}
\caption{Distribution  of seeing  FWHM values  for all  the $grizYJHK$
  frames (from  left to right and  top to bottom).   The median value,
  $\mu$, of each distribution is marked by the dashed line in each
  panel.   Both the value  of $\mu$  and the  width, $\sigma$,  of the
  distributions  are   reported  in  the  top-right   corner  of  each
  plot. \label{seeing}}
\end{center}
\end{figure}
\begin{figure}
\begin{center}
\includegraphics[height=85mm]{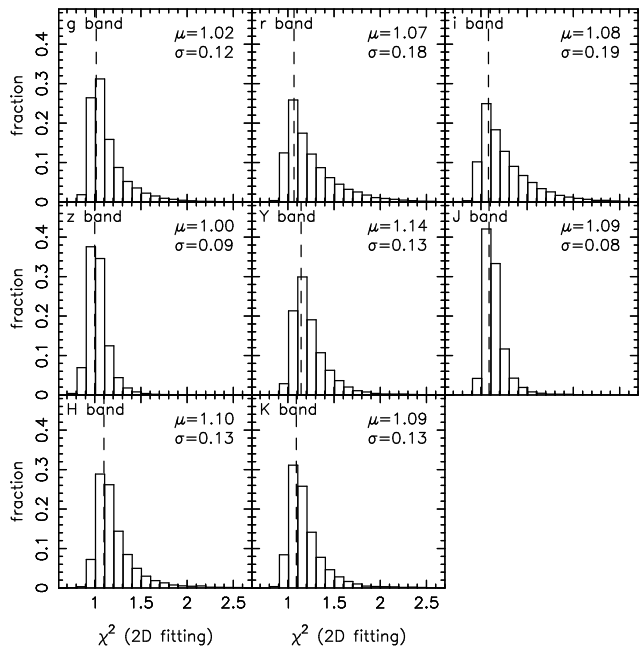}
\caption{Same as Fig.~\ref{seeing} for the $\chi^2$ distributions.
~\label{chi2}}
\end{center}
\end{figure} 

\begin{figure}
\begin{center}
\includegraphics[height=95mm]{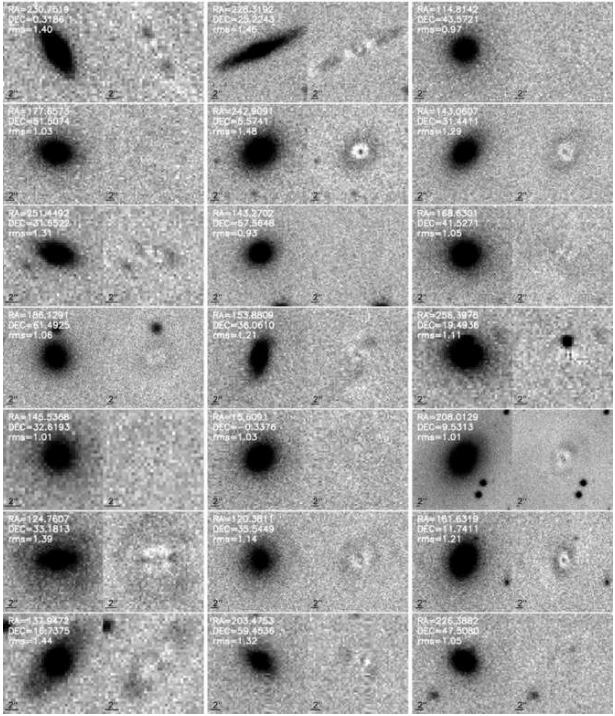}
\caption{Two-dimensional  fit  results  for  galaxies in  r-band  with
  typical $\chi^2$ value ($\chi^2 <  1.5$). Each plot shows the galaxy
  stamp (left)  and the residual map (right)  after model subtraction,
  using the same gray scale of intensity levels.  The spatial scale is
  shown in the  bottom-left corner of the left  plots. For each galaxy
  image, the corresponding celestial coordinates (right ascension, RA,
  and declination, $DEC$, in  degrees) and $\chi^2$ value are reported
  in the upper-left corner.
\label{2DFIT1}}
\end{center}
\end{figure}

\begin{figure}
\begin{center}
\includegraphics[height=95mm]{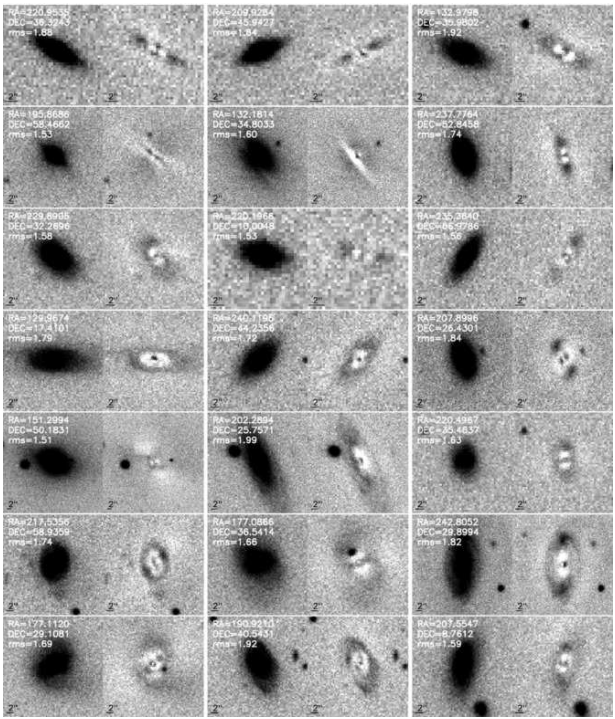}
\caption{Same as Fig.~\ref{2DFIT1} but  for galaxies with high $\chi^2$
  value ($1.5< \chi^2 <2.0$ in r-band).\label{2DFIT2}}
\end{center}
\end{figure}

Fig.~\ref{chi2} compares the  distribution of $\chi^2$ values obtained
from the two-dimensional  fitting of galaxies in each  band.  The peak
and width  values of the  distributions are computed by  the bi-weight
statistics and  reported in  the figure. The  $\chi^2$ is  computed as
follows. For  each galaxy, we  select only pixels 1$\sigma$  above the
local  sky background  value. The  intensity  value of  each pixel  is
computed from  the two-dimensional seeing-convolved  Sersic model. For
the selected pixels,  we compute the $\chi^2$ as  the rms of residuals
between the  galaxy image and  the model. Residuals are  normalized to
the expected noise  in each pixel, accounting for  both background and
photon  noise.   Notice that  this  $\chi^2$  computation is  somewhat
different from  that of  the two-dimensional fitting  procedure, where
the  sum  of square  residuals  over all  the  galaxy  stamp image  is
minimized  (see LdC08).   This explains  the  fact that  all the  peak
values  in  Fig.~\ref{chi2} are  slightly  larger  than  one.  An  eye
inspection of the residual maps, obtained by subtracting the models to
the galaxy  images, shows that  the above $\chi^2$ estimate  is better
correlated   to   the  presence   of   faint  morphological   features
(e.g. spiral  arms, disk, etc...), that  are not accounted  for by the
two-dimensional   model.    This   is  shown   in   Figs.~\ref{2DFIT1}
and~\ref{2DFIT2}.   Both figures  show  residual maps  in the  r-band.
Fig.~\ref{2DFIT1} displays  cases where the  $\chi^2$ is close  to the
peak value  ($\chi^2 <  1.5$), while Fig.~\ref{2DFIT2}  exhibits cases
with higher  $\chi^2$ value ($1.5<  \chi^2 <2.0$).  In most  cases, as
the $\chi^2$ value increases, we can see some faint features to appear
in the residual  maps.  We found that the  percentage of galaxies with
$\chi^2>1.5$  is  not  negligible,   amounting  to  $\sim  17  \%$  in
r-band. Most of  the morphological features are expected  to be caused
by young  stellar populations, hence  disappearing when moving  to NIR
bands,  where the  galaxy light  is  dominated by  the old,  quiescent
stars.   From Fig.~\ref{chi2},  one can  actually see  that  NIR bands
exhibit  a  less pronounced  tail  of  positive  $\chi^2$ values  with
respect   to  the   optical.   {   This   is  also   confirmed  by   a
  Kolmogorov-Smirnov (KS) test.  For instance,  in the case of r and K
  bands, the  KS test gives a  probability smaller than  $1\%$ for the
  corresponding  $\chi^2$  distributions to  be  drawn  from the  same
  parent distribution.}


\subsection{Uncertainty on structural parameters}
\label{sec:err_struc_par}
We estimate  the uncertainties  on structural parameters  by comparing
the  differences  in  \lre,   \mie,  and~\lsn  \,  between  contiguous
wavebands.  To  obtain independent  estimates of the  uncertainties on
SDSS and  UKIDSS parameters, the  comparison is performed for  the $r$
and $i$ bands, and the  $J$ and $H$ bands, respectively. The variation
of \lre, \mie, and \, \lsn \, with waveband depends on the measurement
errors on structural parameters as  well as on the intrinsic variation
of stellar population properties (e.g. age and metallicity) across the
galaxy, which implies a change  of the light profile with waveband. As
it  is well known,  this change  is responsible  for the  existence of
radial  color gradients  inside  ETGs~\citep{Peletier:90}.  The  basic
assumption here is that the $r$  and $i$ ($J$ and $H$) bands are close
enough that  the variation of galaxy  properties from one  band to the
other is dominated  by the measurement errors. Tab.~\ref{tab:dcoldzdt}
shows that this  is the case. In this table  we report the sensitivity
of  color  indices  with   respect  to  age  and  metallicity  between
contiguous wavebands for a Simple Stellar Population (SSP) model.  The
sensitivities to  age and  metallicities, indicated as  $\Delta_t$ and
$\Delta_Z$, are defined  as the derivatives of the  color indices with
respect to $\log  t$ and $\log Z$,  where $t$ and $Z$ are  the age and
metallicity,   respectively.   The   derivatives   are  estimated   as
in~\citet{LdC09}. We use an SSP from the \citet{BrC03} synthesis code,
with  solar  metallicity,  Scalo  IMF,  and an  age  of  $t=10.6$~Gyr,
corresponding  to  a formation  redshift  of  $z_f=3$  in the  adopted
cosmology.  For the SDSS  wavebands, the sensitivities reach a minimum
in $r \! -  \! i$, while they have a maximum  in $g-r$, as expected by
the fact that  this color encompasses the $4000 \rm  \AA$ break in the
spectrum  of  ETGs  at  the  median  redshift  of  the  SPIDER  sample
($z=0.0725$).  The values of $\Delta_t$  and $\Delta_Z$ can be used to
estimate  the  expected  intrinsic  waveband variation  of  structural
parameters.   As  shown by~\citet{SPF09},  bright  ETGs  have a  large
dispersion  in their  radial metallicity  gradients,  $\nabla_Z$, with
$\nabla_Z$ varying  in the range of  about $0$ to  $-0.6$~dex.  On the
contrary,  age  gradients   play  a  minor  role~\citep{LdC09}.   Even
considering a dispersion of  $0.3$~dex in the metallicity gradients of
ETGs, from Tab.~\ref{tab:dcoldzdt}, one can see that the corresponding
scatter in  the $r-i$  internal color gradients  would be  only $0.079
\cdot   0.3  \sim  0.024$mag.    Following~\citet{SpJ93},  for   a  de
Vaucouleurs  profile,  this  implies   an  intrinsic  scatter  in  the
difference of  r- and  i-band effective radii  of only $\sim  2.4 \%$,
hence much  smaller than the  typical measurement error on  $r_e$ (see
below).  For the UKIDSS wavebands, the lowest sensitivities to age and
metallicity are obtained  in $J \!  - \!  H$,  being even smaller than
those of the optical colors.

\begin{table}
%\centering
\small
\begin{minipage}{50mm}
\caption{Sensitivity of color indices of an SSP model to age and metallicity.}
\begin{tabular}{c|c|c}
\hline
color index       &  $\Delta_Z$ & $\Delta_t$  \\
\hline
 g-r &     0.313 &     0.268 \\
 r-i &     0.079 &     0.108 \\
 i-z &     0.107 &     0.131 \\
 z-Y &     0.238 &     0.116 \\
 Y-J &     0.239 &     0.076 \\
 J-H &     0.017 &     0.030 \\
 H-K &     0.134 &     0.049 \\
\hline
  \end{tabular}
\label{tab:dcoldzdt}
\end{minipage}
\end{table}

\begin{figure*}
\begin{center}
\includegraphics[width=110mm]{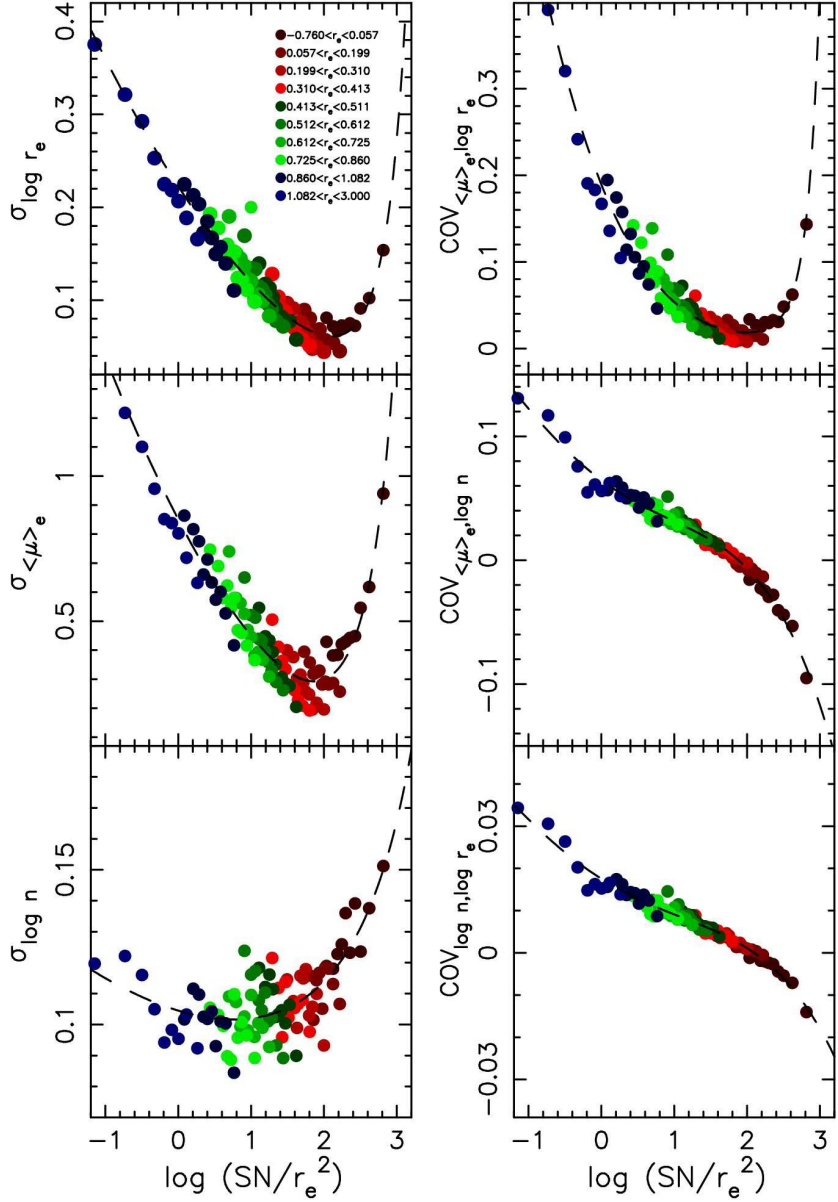}
\caption{Errors on  optical effective parameters as a  function of the
  logarithm  of the  signal-to-noise ($SN$)  per pixel.   In  the left
  panels, from top to bottom, we plot the uncertainty on the effective
  radius, the effective mean surface brightness, and the Sersic index.
  The left  panels, from top to  bottom, plot the  covariance terms of
  the uncertainties  on \lre,  \mie, and~\lsn.  Different  colors show
  different bins  of the effective  radius as shown in  upper-right of
  the top-left panel.  For a  given color, the points shows the values
  of the uncertainties and  the covariance terms obtained in different
  bins of the logarithm of the  $SN$ per pixel.  The dashed curves are
  the best-fit  functional forms used  to model the dependence  of the
  uncertainties on $SN$ (see the text). \label{err_spar}}
\end{center}
\end{figure*} 

\begin{figure*}
\begin{center}
\includegraphics[width=110mm]{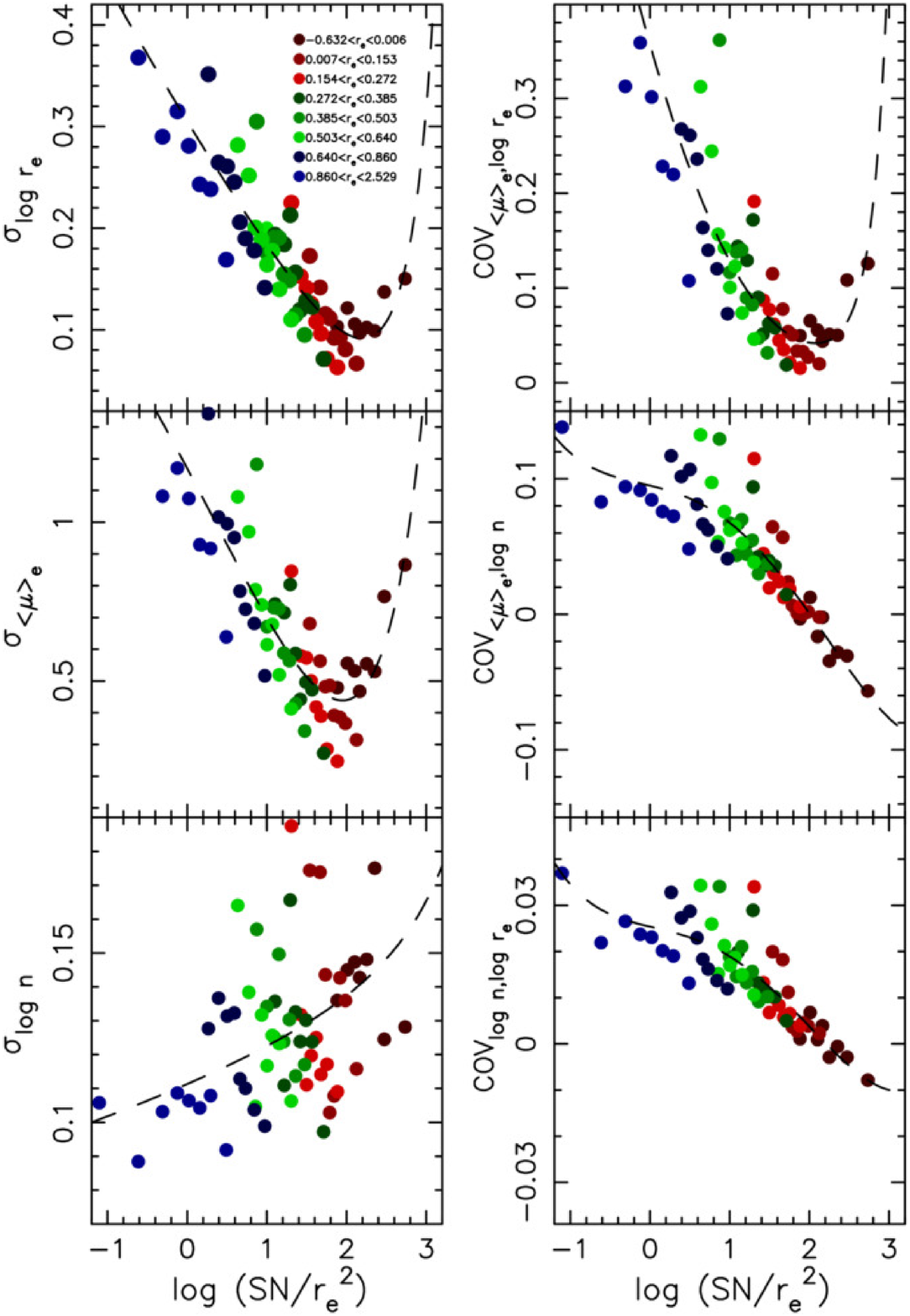}
\caption{Same  as  Fig.~\ref{err_spar}  but  comparing J  and  H  band
  structural parameters.~\label{err_spar_jh}}
\end{center}
\end{figure*} 

The errors on  structural parameters are expected to  be mainly driven
by two parameters, the FWHM $+$ pixel scale of the image (with respect
to the galaxy size) and the signal-to-noise ratio, $SN$. Consequently,
we bin the differences in \lre,  \mie, and \lsn \, between $r$ and $i$
($J$  and  $H$)  bands with  respect  to  the  logarithm of  the  mean
effective  radius, \lre,  and the  $SN$ per  unit area  of  the galaxy
image, $SN/r_e^2$. The $SN$ is  defined as the mean value, between the
two   bands,   of  the   inverse   of   the   uncertainties  on   Kron
magnitudes. Each bin is chosen to have the same number of galaxies. In
a given  bin, we  estimate the measurement  errors on \lre,  \mie, and
\lsn  \,  from  the  mean  absolute  deviation  of  the  corresponding
differences in that bin. We  refer to the measurement uncertainties as
\slre, \smie,  and \sln, respectively.  Since the  errors on effective
parameters are  strongly correlated, we also  derive the corresponding
covariance terms, $COV_{Y,X}$, in each  bin, where $X$ and $Y$ are two
of the three quantities \lre, \mie, and \lsn. For each pair of $X$ and
$Y$, we perform a robust linear fit~\footnote{The robust regression is
  performed by  minimizing the  sum of absolute  residuals of  the $Y$
  vs. the $X$ differences.} of the differences in $Y$ as a function of
the  corresponding differences  in  $X$. The  quantity $COV_{Y,X}$  is
obtained from the  slope, $s$, of the fitted  relation as $COV_{Y,X}=s
\times \sigma_X^2$,  with the constraints  $|COV_{Y,X}| \le \sigma_{Y}
\times  \sigma_{X}$,  and $COV_{Y,X}=COV_{X,Y}$.   Fig.~\ref{err_spar}
plots the quantities \slre, \smie, and \sln \, as well as the relevant
covariance terms  as a function  of $\log(SN/r_e^2)$, for the  $r$ and
$i$ bands.  The errors on effective parameters are strongly correlated
to the $SN$  per unit area.  For $\log(SN/r_e^2)  \widetilde{<} 2$, as
the signal-to-noise per unit area decreases, the errors tend to become
larger.   For  $\log(SN/r_e^2)  \widetilde{>}  2$, all  galaxies  have
effective radii comparable or even  smaller than the pixel scale, with
large  values of $\log(SN/r_e^2)$  corresponding to  smaller effective
radii. As  a result,  the error on  the effective parameters  tends to
increase as well.  The \sln \,  exhibits a similar behavior to that of
\slre  \,  and  \smie,  though  with  a  larger  dispersion  at  given
$\log(SN/r_e^2)$.   For  the errors  on  effective parameters  (\slre,
\smie, and \slremie) and Sersic  index (\sln), the trends exhibited in
Fig.~\ref{err_spar}  are well  described by  the  following, empirical
functional form:
\begin{equation}
y=P_2(x)+\frac{|c_1|}{(x-5)^{c_2}},
\label{eq:errors}
\end{equation}
where $y$  is one of the  quantities \slre, \smie, and  \slremie, $x =
\log(SN/r_e^2)$, $P_2$  is a  second order polynomial  function, while
$c_1$  and $c_2$  are two  parameters describing  the increase  of the
error values at  high $SN$ per unit area.   For the quantities \smieln
\, and \slnlre, the trends  in Fig.~\ref{err_spar} can be modeled by a
fourth  order polynomial  function. We  have fitted  the corresponding
functional  forms  by minimizing  the  sum  of  absolute residuals  in
$y$. The best-fit curves are exhibited in Fig.~\ref{err_spar}.

Fig.~\ref{err_spar_jh}  plots the errors  on structural  parameters as
derived with the above procedure for the $J$ and $H$ bands. The trends
are similar to  those obtained for the optical  parameters, but with a
larger  dispersion, at  given  $SN$  per unit  area,  which is  likely
explained by  the fact that  the number of  galaxies in each  bin with
available  photometry  in $J$  and  $H$ is  smaller  (by  a factor  of
$\sim8$)  than that  in $r$  and $i$.   The uncertainties  on  the NIR
parameters  are on  average larger  than  those in  the optical.   For
instance, at $\log(SN/r_e^2) = 2$, the uncertainty on \lre \, is $\sim
0.06$~dex in the optical, increasing to $\sim 0.1$~dex in the NIR. The
median errors  on \lre,  \mie, and \lsn,  amount to $\sim  0.1$, $\sim
0.5$~$mag/arcsec^2$, and $\sim 0.1$, respectively, in the optical, and
to $\sim 0.15$, $\sim 0.6$~$mag/arcsec^2$, and $\sim 0.12$ in the NIR.
This difference  can be qualitatively  explained by the fact  that (i)
the  stars  used  for the  PSF  modeling  have,  on average,  a  lower
signal-to-noise ratio  in the NIR than  in the optical,  and (ii) that
the ratio of  galaxy effective radii to the pixel  scale of the images
is  smaller  in   the  NIR  than  in  the   optical.   The  trends  in
Fig.~\ref{err_spar_jh} are  modeled with the same  functional forms as
for the optical data.

We  use  the  above  analysis  to  assign  errors  to  the  structural
parameters to  each galaxy  in the SPIDER  sample, for  each waveband.
For a given galaxy, we first calculate its $SN$ per unit area and then
use the  best-fitting functional forms  to assign \slre,  \smie, \sln,
and the corresponding covariance terms.  For the $gri$ bands, we adopt
the functional forms obtained from the $r-i$ comparison, while for the
$JHK$ bands we  adopt the values obtained from  the comparison of $J$-
and $H$-band parameters.  In the $z$  and $Y$ bands, we apply both the
optical  and NIR  functional  forms,  and then  derive  the errors  by
interpolating the  two error estimates  with respect to  the effective
wavelength of the passbands.

\section{Color-magnitude relations}
\label{sec:CM}
As a  first step  in the  comparison of optical  to NIR  properties of
ETGs,  we  start  to  analyze  the  differences  in  their  integrated
properties,  i.e.  the  color  indices. The  goal  is comparing  total
Sersic  magnitudes  in  the  different  wavebands,  in  order  to  (i)
characterize the completeness of the SPIDER sample in the space of the
(Sersic) effective parameters  (Sec.~\ref{sec:compl}), and (ii) select
suitable samples of ETGs for the analysis of the FP (see papers II and
III).   Thus, we  estimate the  color indices  using the  Sersic total
magnitudes,  rather than aperture  magnitudes as  in most  of previous
studies, and refer to them also as the {\it total} galaxy colors. This
is by  itself an important  issue that will  be addressed in  a future
contribution dealing  with the different ways of  measuring colors (de
Carvalho et al. 2010, in  preparation) which certainly goes beyond the
scope of this paper.

\begin{figure*}
\begin{center}
\includegraphics[height=80mm]{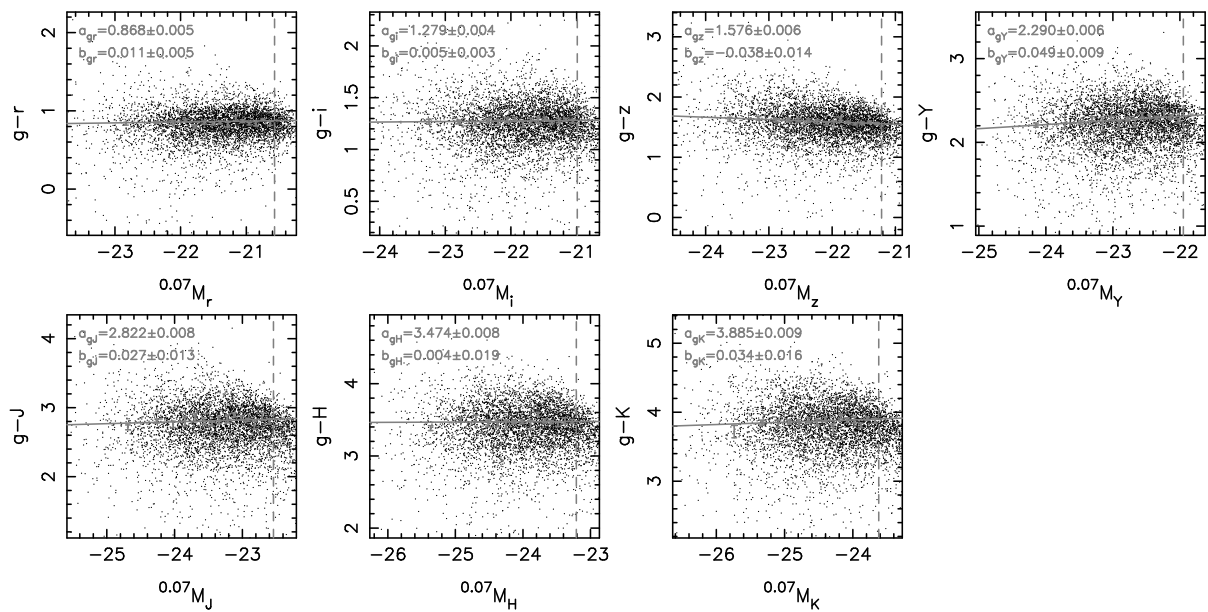}
\caption{Color--magnitude relations for  all the $5,080$ galaxies with
  photometry available  in $grizYJHK$. From  left to right and  top to
  bottom, galaxy  colors in the  form of $g-X$, with  $Z=rizYJHK$, are
  considered. For each panel: the  slope and offset of the CM relation
  are  reported  in  the  upper-left corner,  with  the  corresponding
  1$\sigma$ standard uncertainties; the colored circles mark the peak
  of the color distribution  in magnitude bins, with the corresponding
  1$\sigma$  error bars;  the  solid line  shows  the best-fitting  CM
  relation; the dashed line mark  the magnitude cut ion that band that
  corresponds  to a  magnitude limit  of $-20.55$  in the  r-band (see
  text).~\label{fig:CM}}
\end{center}
\end{figure*} 

The  comparison of  color  indices is  performed  by constructing  the
color--magnitude diagrams for different pairs of wavebands.  Since the
ETG sample has photometry available  in eight wavebands, we can derive
seven different color--magnitude relations.  We consider galaxy colors
in  the  form  of $g-X$  with  $X=rizYJHK$,  and  we write  the  color
magnitude relations as:
\begin{equation}
 g-X = a_{_{gX}}+b_{_{gX}} X,
\label{eq:CM1}
\end{equation}
where $a_{_{gX}}$  and $b_{_{gX}}$ are  the offsets and slopes  of the
relations.   To simplify  the notations,  we also  set  $a_{gg}=0$ and
$b_{_{gg}}=0$.  Fig.~\ref{fig:CM}  plots the  CM diagrams for  all the
$5,080$ ETGs with available data  in all wavebands. In order to derive
$a_{_{gX}}$ and $b_{_{gX}}$,  we first bin each $g-X$  vs. $X$ diagram
with respect  to the magnitude $X$.   We adopt $N=15$  bins, with each
bin including the same number  of galaxies. Varying the number of bins
in the  range of $15$ to $25$  changes the slope and  offset values by
less than 1$\%$. For each bin, we derive the peak of the corresponding
distribution  of  galaxy colors  by  applying  the bi-weight  location
estimator~\citep{Beers:90}.   The  uncertainty of  the  peak value  is
estimated as  the standard  deviation of the  peak values  obtained in
$1000$ bootstrap iterations. This procedure has the advantage of being
insensitive  to outliers in  the color  distribution.  The  binning is
performed up  to a magnitude  limit $X$, obtained by  transforming the
2DPHOT r-band completeness  limit of $-20.55$ (Sec.~\ref{sec:compl_r})
through the  median values of the  $g-X$ total colors.   The values of
$a_{_{gX}}$  and $b_{_{gX}}$ are  then derived  by fitting  the binned
values  of $g-X$  vs.   $X$, with  an  ordinary least-squares  fitting
procedure  with $g-X$  as  dependent variable.   The uncertainties  on
$a_{_{gX}}$  and   $b_{_{gX}}$  are  obtained   by  randomly  shifting
($N=1000$  times)  the  binned  values  of $g-X$  according  to  their
uncertainties. The values of $a_{_{gX}}$ and $b_{_{gX}}$ are exhibited
in Fig.~\ref{fig:CM}.   It is interesting to notice  that, using total
colors,  the  CM  relations  are  essentially flat,  with  the  slopes
$b_{_{gX}}$ being mostly consistent with zero within the corresponding
uncertainties. In  particular, the value of  $b_{_{gX}}$ is consistent
with zero  at less than $2\sigma$  in $g-r$, $g-i$,  $g-J$, $g-H$, and
$g-K$,  while  in $g-z$  and  $g-Y$ the  slopes  differ  from zero  at
$\sim3\sigma$ and  $\sim5\sigma$ levels, respectively. We  do not find
here any systematic  steepening of the CM relation  when enlarging the
waveband baseline from $g-r$ to  $g-K$, as expected if the {\it total}
CM relation would be purely  driven by a mass--metallicity relation in
ETGs.  Following~\citet{Scodeggio:01}, we  can explain this surprising
result by the fact that we use total color indices. ETGs have negative
color  gradients, with color  indices becoming  bluer from  the galaxy
center to its periphery. As a  result, when adopting colors in a fixed
aperture, one  is measuring  the color inside  a smaller  region (with
respect to  \re) for  the brightest (hence  larger) galaxies  than for
faintest galaxies in the sample. This leads to a misleading steepening
of  the  CM relation.   We  notice  that  using de  Vaucouleurs  model
magnitudes from  SDSS would  essentially lead to  the same  effect, as
model magnitudes  are estimated in a  fixed aperture for  all the SDSS
wavebands (see~\citealt{EDR}), and the SDSS effective radii tend to be
more underestimated  (with respect to  the Sersic $r_e$)  for brighter
than   for  fainter   galaxies   (see  Fig.~\ref{fig:CONF_SPAR_n}   of
Sec.~\ref{sec:conf_SDSS_2DPHOT}).   Since the values  of $b_{_{gX}}=0$
are mostly consistent  with zero and our aim here  is that of relating
total magnitudes  among different wavebands (rather  than performing a
detailed study of CM relations), we have decided to set $b_{_{gX}}=0$,
and derive $a_{_{g-X}}$ as the median  of the $g-X$ peak values in the
different magnitude bins. {  The values of $a_{_{g-X}}$, together with
  the      corresponding     uncertainties,     are      listed     in
  Tab.~\ref{tab:meancol}.   The uncertainties  are the  errors  on the
  median values. They are estimated from the width of the distribution
  of median values for $200$ bootstrap iterations.}

Fixing the limiting magnitude of the ETG's sample in a given band, $W$
($W=grizYJHK$), one  can use the CM  relations to map  that limit into
{\it  equivalent} magnitude limits,  $X_{lim}$ ($X=grizYJHK$),  in all
wavebands. From Eq. 3, one obtains:
\begin{equation}
 X_{lim} =
\frac{a_{_{gW}}-a_{_{gX}}}{1+b_{_{gX}}}+\frac{1+b_{_{gW}}}{1+b_{_{gX}}}
W_{lim}.
\label{eq:CM}
\end{equation}
In the  particular case of  $b_{_{gX}}=0$, one obtains  the simplified
expressions, $X_{lim} = a_{_{gY}}-a_{_{gX}}+Y$, which is used in paper
II to  analyze the  FP relation for  {\it color--selected}  samples of
ETGs.

%In practise, we adopted an r-band magnitude limit of $r=-20.6$, and
%derived the corresponding values of $X_{lim}$ from Eqs.~\ref{eq:CM_b0}.
%The values of $X_{lim}$, obtained with this procedure, are those reported
%in Tab.~\ref{tab:FP_samples}. The values of $r=-20.6$ was chosen to ensure
%that for ech band the value of $X_{lim}$ is fainter than the completeness
%magnitude in that band.

\begin{table}
%\centering
\small
\begin{minipage}{50mm}
 \caption{Median peak values, $a_{g-X}$, of the distributions of total
colors, $g-X$ ($X=rizYJHK$). The uncertainties are 1$\sigma$ standard errors
{ on the median values}.}
  \begin{tabular}{c|c}
   \hline
color       &   peak value    \\
\hline
$g-r$  &   $0.852 \pm 0.007$ \\
$g-i$  &   $1.274 \pm 0.004$ \\
$g-z$  &   $1.589 \pm 0.008$ \\
$g-Y$  &   $2.282 \pm 0.012$ \\
$g-J$  &   $2.806 \pm 0.012$ \\
$g-H$  &   $3.472 \pm 0.012$ \\
$g-K$  &   $3.854 \pm 0.018$ \\
\hline
  \end{tabular}
\label{tab:meancol}
\end{minipage}
\end{table}

\section{Distribution of structural parameters from \MakeLowercase{g} through $K$}
\label{sec:dist_par}

\begin{figure}
\begin{center}
\includegraphics[height=85mm]{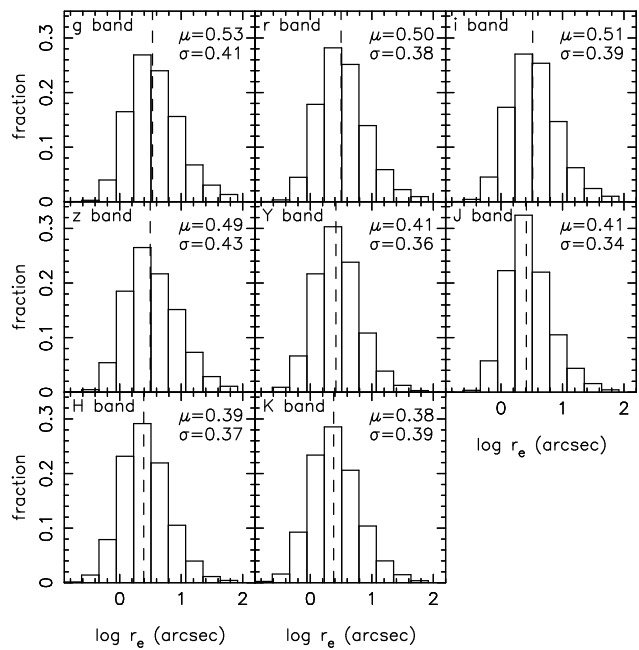}
\caption{Distributions of \lre \, for all the $grizYJHK$ wavebands
  (from left to right and top to bottom). The median value, $\mu$, of
  each distribution is marked by the vertical dashed line. The $\mu$
  \, and width values (see the text) are reported in the upper--right
  corner of each panel.~\label{logre} }
\end{center}
\end{figure}

\begin{figure}
\begin{center}
\includegraphics[height=85mm]{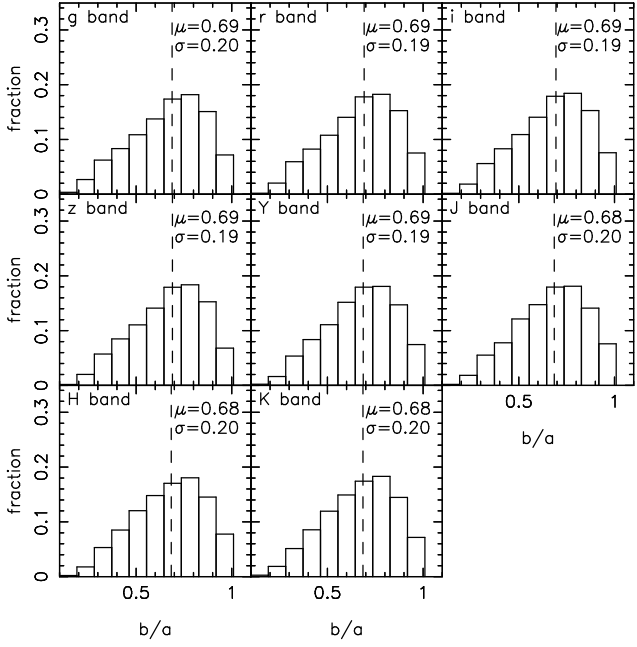}
\caption{Same  as  Fig.~\ref{logre}  for  the  distribution  of  axis
  ratios. \label{ba}}
\end{center}
\end{figure}

\begin{figure}
\begin{center}
\includegraphics[height=85mm]{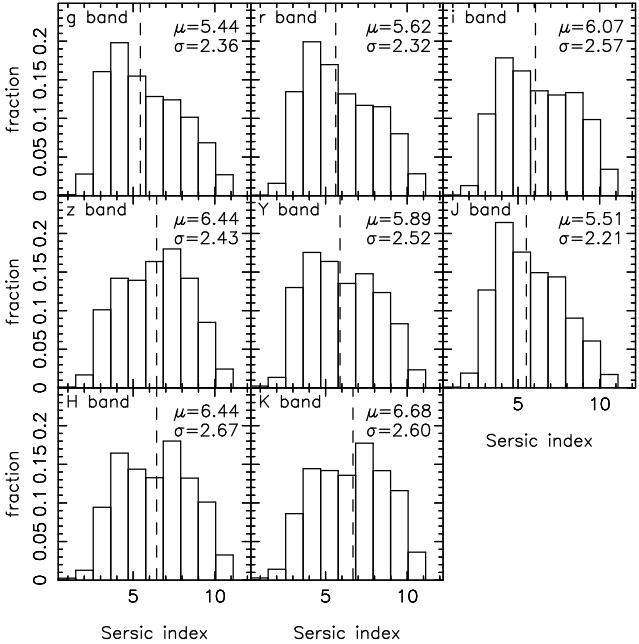}
\caption{Same as Fig.~\ref{logre} for the distribution of Sersic indices.
\label{sersic}}
\end{center}
\end{figure}

Figures~\ref{logre}  to~\ref{sersic}   exhibit  the  distributions  of
2DPHOT  Sersic parameters  from $g$  through  $K$. For  each band,  we
select   all    the   galaxies    available   in   that    band   (see
Sec.~\ref{sec:DATA}).   Each  distribution  is  characterized  by  its
median  value,  $\mu$,  and  the  width, $\sigma$,  estimated  by  the
bi-weight   statistics~\citep{Beers:90}.   Both   values,   $\mu$  and
$\sigma$, are reported in the plots.

Fig.~\ref{logre}  compares the distributions  of effective  radii. The
most noticeable feature is that  the median value of \lre \, decreases
smoothly from  the optical  to the NIR,  varying from  $\sim 0.53$~dex
($r_e  \sim  3.4''$) in  $g$  to $\sim  0.38$~dex  in  $K$ ($r_e  \sim
2.4''$).  This change  of $\sim  0.15$~dex corresponds  to  a relative
variation of $\sim 35\%$ in \re, and is due to the fact that ETGs have
negative  internal color  gradients, with  the light  profile becoming
more concentrated  in the center as  one moves from  shorter to longer
wavelengths.  Had  we  used  the  peak values  of  the  distributions,
estimated with  the bi-weight  statistics, rather than  median values,
the relative  variation in \re \,  would have been  $\sim 31\%$ ($\sim
0.13$~dex)  instead of $\sim  35 \%$.  The optical--NIR  difference in
\lre  \,  is   in  agreement  with  the  value   of  $\sim  0.125$~dex
($\sim29\%$) reported  by~\citet{KoIm:05} for  the sample of  273 ETGs
from~\citet{Pahre:99},  between the  $V$ and  $K$ bands.   Notice also
that  the   relative  change  in  \re  \,   fortuitously  matches  the
improvement in  average seeing FWHM  between the g- and  K-band images
(Sec.~\ref{sec:struc_par}), which makes  the measurement of structural
parameters   across   the  SPIDER   wavelength   baseline  even   more
homogeneous.

Fig.~\ref{ba} compares the distributions  of values of the axis ratio,
$b/a$, of the  best-fitting Sersic models.  The median  as well as the
width values  of $b/a$  turn out to  be essentially constant  from $g$
through $K$, amounting to $\sim 0.7$ and $\sim 0.2$, respectively. The
consistency of the $b/a$  distributions with wavebands is in agreement
with that found by~\citet{HyB:09a}  when comparing $g$- and $r$- bands
$b/a$ values  from SDSS. As  expected, the fraction of  ETGs decreases
dramatically  at  low values  of  $b/a$ with  only  a  few percent  of
galaxies having axis ratios as low as $0.3$.

Fig.~\ref{sersic}  compares  the distributions  of  the Sersic  index,
$n$. Since we have selected bulge-dominated galaxies ($fracDev_r>0.8$,
see Sec.~\ref{sec:DATA}), all the objects exhibit a Sersic index value
larger than one,  i.e. no galaxy has an  exponential (disk-like) light
profile.  In  particular, the  fraction of ETGs  becomes significantly
larger  than zero  above the  value of  $n \sim  2$,  which, according
to~\citet{Blanton:03b},  roughly corresponds  to the  separation limit
between blue and  red galaxies in the SDSS.   The distributions show a
large  scatter, with  $n$ ranging  from $\sim  2$ to  $\sim  10$.  The
median value of $n$ is around $6$ for all wavebands, without any sharp
wavelength dependence. On the other  hand, we see some marginal change
in the shape of the  distribution with waveband.  In the optical $gri$
wavebands, a  peak in the distribution  is evident around  $n \sim 4$.
The  distributions   become  essentially  flat  in   the  other  (NIR)
wavebands, with the exception of $J$  where a peak is still present at
$n \sim 4$.  Some caution goes in interpreting these changes in shape.
First, notice  that observations and  data reduction in the  J-band of
UKIDSS-LAS are carried out in a somewhat different manner with respect
to  the other  wavebands (see  Warren et  al.2007).   A micro-stepping
procedure, with  integer pixel  offsets between dithered  exposures is
performed.   Images  are  interleaved  to  a subpixel  grid  and  then
stacked.   This procedure  results  in a  better  image resolution  of
$0.2''/pixel$,  with a  better  accuracy of  the astrometric  solution
(useful for proper motion's measurements). We cannot exclude that this
difference in  data reduction affects  the J-band distribution  of $n$
values.    Moreover,    as   seen   in   Sec.~\ref{sec:err_struc_par},
uncertainties on  structural parameters  change from $g$  through $K$,
hence  preventing a  straightforward comparison  of the  shape  of the
distributions among different wavebands.

\begin{figure}
\begin{center}
\includegraphics[width=80mm]{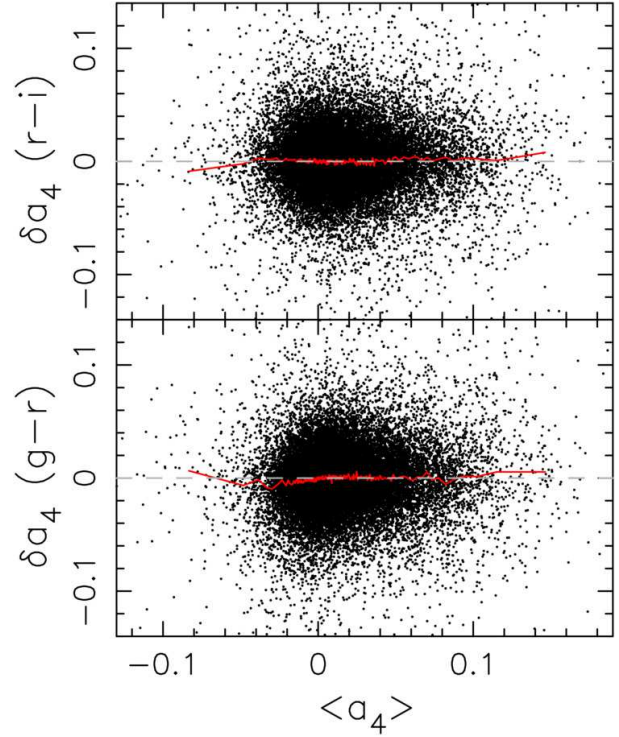}
\caption{Differences  of $a_4$ estimates  between $r-i$  (upper panel)
  and $g-r$  (lower panel)  as a function  of the median  $a_4$ value,
  $<a_4>$,  in the  $gri$ wavebands.  The blue  dashed line  marks the
  value of zero. The red curves  are obtained by binning the data with
  respect to $<a_4>$.~\label{conf_a4}}
\end{center}
\end{figure} 

In order to analyze the waveband dependence of the $a_4$ parameter, we
compare the $a_4$ values among  contiguous wavebands for the sample of
39,993  ETGs.   The  comparison  is  performed using  only  the  $gri$
wavebands,  where  the $a_4$  estimates  are  derived (see  Sec.~3.2).
Fig.~\ref{conf_a4}  plots  the  differences  in $a_4$,  $\delta  a_4$,
between $g$  and $r$,  and $r$ and  $i$, as  a function of  the median
$a_4$ value, $<a_4>$. The $\delta  a_4$ values are binned with respect
to  $<a_4>$, with  each bin  including  the same  number ($N=200$)  of
galaxies.   For a  given bin,  the median  difference of  $\delta a_4$
values is  computed.  The  median values are  plotted as  a continuous
curve in Fig.~\ref{conf_a4}, showing that there is no systematic trend
of $a_4$ from $g$ through $i$.

\section{Comparison of SDSS and 2DPHOT structural parameters}
\label{sec:conf_SDSS_2DPHOT}

We compare  the effective parameters  measured with 2DPHOT  with those
derived from  the SDSS photometric  pipeline \photo, that  fits galaxy
images   with   two-dimensional   seeing  convolved   de   Vaucouleurs
models~\citep{EDR}.  From  now on, the  differences are always  in the
sense of  SDSS$-$2DPHOT.  The comparison  is done in r-band,  by using
the      entire      SPIDER       sample      of      39,993      ETGs
(Sec.~\ref{sec:DATA}). Effective radii along the galaxy major axis are
retrieved  from  the  SDSS   archive,  and  transformed  to  equivalent
(circularized)  effective radii,  $r_e$,  with the  axis ratio  values
listed in SDSS.  The effective mean surface brightness,  \mie, is then
computed from the circularized  effective radii and the de Vaucouleurs
model  magnitude, \mdevr, using  the definition  \mie$=$\mdevr$+5 \log
(r_e)+2.5\log(2  \pi)$.  All magnitudes  are  dereddened for  galactic
extinction  and k-corrected  (see  Sec.~\ref{sec:int_phot}). For  both
2DPHOT and  \photo, we denote the  effective parameters as  \re \, and
\mie,  saying explicitly  when we  refer to  either one  or  the other
source.

\begin{figure}
\begin{center}
\includegraphics[height=85mm]{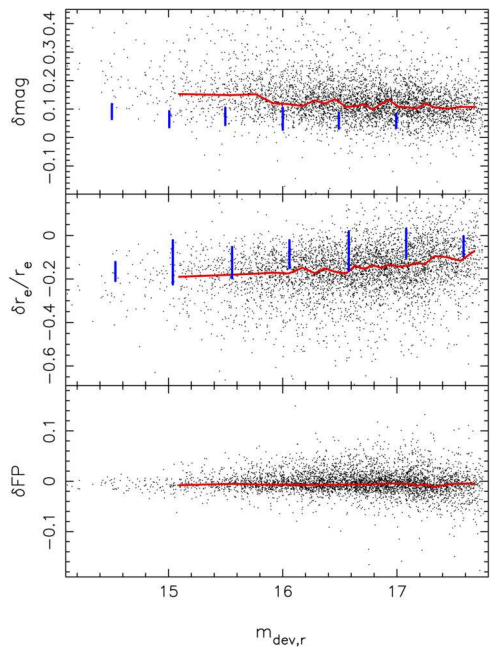}
\caption{Differences  in  total  magnitude,  $\delta  \,  mag$  (upper
  panel), effective radius, $\delta  r_e/r_e$ (middle panel), and {\it
    FP} parameter, $\delta FP$ (lower panel), between SDSS and 2DPHOT,
  as a function  of the SDSS r-band model  magnitude. All values refer
  to the  r-band. Red curves  are obtained by binning  the differences
  with respect to \mdevr (see the  text). The blue bars mark the range
  of expected differences due to the sky overestimation effect present
  in the  SDSS parameters  estimation. The size  of the blue  bars has
  been  estimated  from fig.~6  of~\citet{DR7}  for  $\delta mag$  and
  fig.~3             of~\citet{DR6}             for            $\delta
  r_e/r_e$.~\label{fig:CONF_SPAR_n4}}
\end{center}
\end{figure}

In  order to  compare the  method  itself to  derive \re  \, and  \mie
(rather than  the kind of model,  i.e. Sersic vs.  de Vaucouleurs), we
start  by comparing  the effective  parameters of  galaxies  for which
2DPHOT gives a  Sersic index of $n \sim 4$. To  this effect, we select
all the ETGs with $n$ in the range of $3.7$ to $4.3$, considering only
galaxies  with  better quality  images  (seeing  FWHM $<1.5''$).  This
selection    results    into   a    subsample    of   $4,525$    ETGs.
Fig.~\ref{fig:CONF_SPAR_n4}  plots   the  differences  in   the  total
magnitude, effective radius,  and {\it FP} parameter as  a function of
\mdevr, where $FP=$\lre$-0.3$\mie \, is the combination of \lre \, and
\mie \, entering  the FP relation (Saglia et al.  2001).  For the \re,
we normalize the differences to the \photo \, \re \, values.  For each
quantity, the differences are binned  with respect to \mdevr, each bin
including the same  number ($N=200$) of galaxies. In  a given bin, the
peak  value of  the distribution  of  differences (red  curves in  the
figure) is computed by the bi-weight statistics~\citep{Beers:90}.
The SDSS  total magnitudes  and effective radii  differ systematically
from  those obtained  with  2DPHOT, with  total magnitudes  (effective
radii) being fainter  (smaller) with respect to those  of 2DPHOT. This
effect tends to disappear for faint galaxies: the absolute differences
in  total magnitude  decrease  from $\sim  0.15$mag  ($\sim 20\%$)  at
\mdevr$  \sim 15$  to $\sim  0.1$mag  ($6\%$) at  \mdevr$ \sim  17.7$.
These  differences  are in  the  same  sense  as those  reported  from
previous studies (e.g.~Bernardi  et al. 2007, \citealt{Lauer:07}), and
can at  least partly  be accounted by  the sky  overestimation problem
affecting SDSS  model parameters~\citep{DR6, DR7}. \photo  \, tends to
overestimate  the sky  level near  large bright  galaxies,  leading to
underestimate both  total fluxes and  effective radii. The  effect has
been  quantified from  the  SDSS team  by  adding simulated  $r^{1/4}$
seeing-convolved  models to  SDSS images,  and recovering  their input
parameters  through \photo.  Fig.~\ref{fig:CONF_SPAR_n4}  compares the
range of  values for differences  between input and  output parameters
from the  SDSS simulations  (blue bars), with  what we find  here. For
\re,  the  average   \photo-2DPHOT  differences  are  only  marginally
consistent  with those expected  from the  simulations. For  the total
magnitudes,  we  find   larger  systematic  differences,  with  2DPHOT
magnitudes  being brighter,  by  a few  {  tenths} of  mag, than  what
expected from the simulation's  results.  We should notice that larger
differences in magnitude (and perhaps in \re), in the same sense as we
find  here, have  also been  reported by~\citet{DOF08}  when comparing
their effective  parameters with those from SDSS.   Moreover, the SDSS
simulations have  been performed by assuming  a given luminosity--size
relation for ETGs, which might be slightly different for ETG's samples
selected according to different criteria.   Since one main goal of the
SPIDER project is that of analyzing  the FP relation, we have to point
out  that,  although  the  differences  in  \re  \,  and  \mt  \,  are
significant, the  $FP$ quantity is  in remarkable good  agreement when
comparing \photo \,  to 2DPHOT. The average difference  in the $FP$ is
less than  a few percent and does  not depend on the  magnitude.  As a
consequence (see paper  II), the FP coefficients change  by only a few
percent when using either 2DPHOT or \photo \, effective parameters.

\begin{figure}
\begin{center}
\includegraphics[width=80mm]{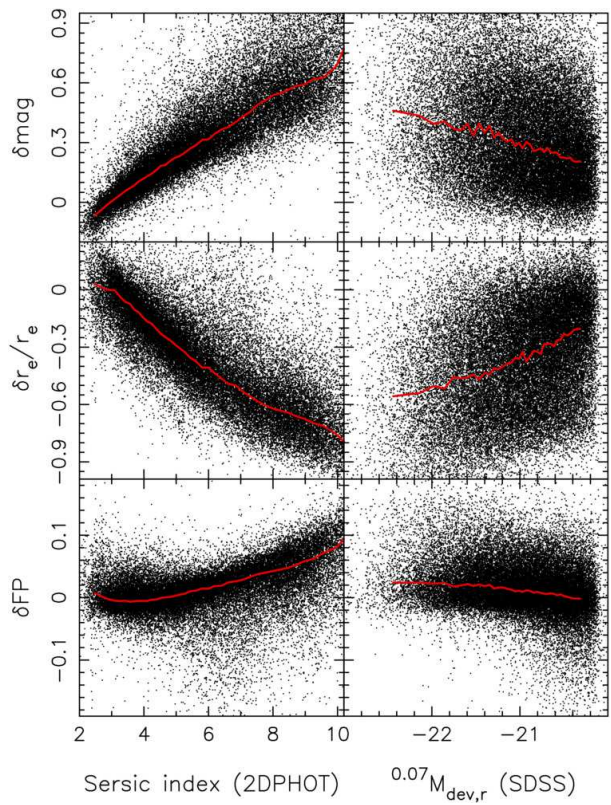}
\caption{Comparison of SDSS and 2DPHOT parameters as a function of the
  2DPHOT  Sersic  index  (left  panels),  and  the  SDSS  total  model
  magnitude (right  panels).  From top to bottom,  the same quantities
  as in  Fig.~\ref{fig:CONF_SPAR_n4} are compared. The  red curves are
  obtained by binning the  data as in Fig.~\ref{fig:CONF_SPAR_n4}, but
  with each bin including $400$ galaxies.  ~\label{fig:CONF_SPAR_n}}
\end{center}
\end{figure}

Fig.~\ref{fig:CONF_SPAR_n} compares differences  between \photo \, and
2DPHOT parameters as  a function of the Sersic index,  $n$, as well as
the  (SDSS) absolute  model magnitude  in r  band, $^{0.07}M_{dev,r}$.
Differences are binned as in  Fig. 15, considering only the $N=39,091$
galaxies  with  better  image  quality (see  above).   The  comparison
reveals large systematic differences, that strongly correlate with the
Sersic index $n$.  As $n$  increases, { Sersic total magnitudes become
  brighter -- while  Sersic effective radii become larger  -- than the
  SDSS  values.  The  former trend  is consistent  with  that reported
  by~\citet{Graham:05}}. Similar, but  weaker, trends are also present
as a  function of  the galaxy magnitude,  when moving from  fainter to
brighter  galaxies.  This  is  somewhat expected,  as  ETGs exhibit  a
luminosity-Sersic  index relation,  with brighter  galaxies  having on
average    {    larger     $n$~\citep{CCD93}}.     The    trends    of
Fig.~\ref{fig:CONF_SPAR_n}    are    similar    to   those    obtained
by~\citet{DOF08} (see  their fig.~1), when  comparing Sersic effective
parameters to  the \photo  \, quantities. As  noticed above,  the $FP$
parameters are more stable with  respect to the fitting procedure than
the other quantities.   In particular, the quantity $FP$  shows only a
weak  dependence  on  galaxy   magnitude  (see  lower-right  panel  of
Fig.~\ref{fig:CONF_SPAR_n}), with  an end-to-end average  variation of
only $0.02$~dex ($\sim  5\%$). { Since the FP can be  seen as a linear
  relation  between  the $FP$  quantity  and  velocity dispersion  (or
  galaxy  magnitude~\footnote{  Notice that  the  Sersic  "n" is  also
    correlated  with  velocity  dispersion~\citep{Graham:02}, but  the
    correlation   exhibits   a   large   dispersion.},   through   the
  Faber--Jackson relation),  the weak dependence of  the $FP$ quantity
  with magnitude implies that the  coefficients of the FP are expected
  not  to  change  significantly  when  using either  2DPHOT  or  SDSS
  parameters (see paper II).}

\section{Spectroscopy}
\label{sec:spectros}

\subsection{Velocity dispersions from SDSS and STARLIGHT}
\label{sec:veldisp}
{ We have re-computed central velocity dispersions for all the ETGs
  in the SPIDER sample.   Velocity dispersions are usually measured by
  comparing  the   observed  galaxy  spectrum   with  single  spectral
  templates,  which  are  assumed  to describe  the  dominant  stellar
  population of the galaxy.  For  ETGs, the spectra of red giant stars
  are usually adopted.  On the  other hand, ETGs frequently show mixed
  stellar  populations,  and   this  might  significantly  affect  the
  $\sigma_0$ estimate  for some fraction of the  ETG's population.  We
  derive the $\sigma_0$'s  with the same procedure as  in the SDSS-DR6
  pipeline,    i.e.   the   direct    fitting   of    galaxy   spectra
  (see~\citealt{DR6}), but instead  of using single spectral templates
  as in  the SDSS pipeline,  we construct a  mixed-population spectral
  template for  each galaxy.   This is done  as part of  the automatic
  procedure   described  in  STARLIGHT~\citep{CID05}   where  velocity
  dispersion   and  stellar   population  parameters   are  determined
  simultaneously.  Hence, the new velocity dispersion values should be
  virtually  unaffected by  the  different kinematics  of the  various
  stellar components.

For  each galaxy,  we run  the  spectral fitting  code STARLIGHT  (Cid
Fernandes  et al.   2005) to  find the  combination of  single stellar
population (SSP)  models that, normalized  and broadened with  a given
sigma, best matches the  observed spectrum (also normalized), which is
first de-redshifted  and corrected for  extinction. We use  SSP models
from the MILES  galaxy spectral library, with a  Salpeter Initial Mass
Function truncated at  lower and upper cutoff mass  values of 0.01 and
120 $\rm M_{\odot}$, respectively~\citep{VAZDEKIS10}. These models are
based on  the MILES stellar  library ~\citep{SANCHEZ06}, which  has an
almost  complete  coverage  of  stellar atmospheric  parameters  at  a
relatively  high and  nearly constant  spectral resolution  of 2.3$\rm
\AA$  (FWHM). This  resolution is  better than  that of  SDSS spectra,
allowing  us to  suitably degrade  the  spectral models  to match  the
resolution of the observed spectra (see below).}
\begin{figure}
\begin{center}
\includegraphics[height=80mm]{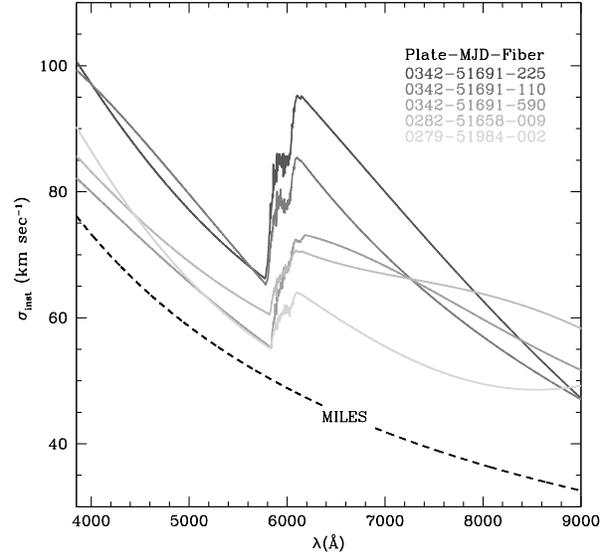}
\caption{Wavelength  dependent instrumental  resolution  for the  SDSS
  spectra of  five randomly selected  ETGs. The resolution  values are
  taken from the  fits spectral files in the  SDSS archive. The dashed
  line shows the resolution  variation for the MILES SSP-models, which
  have a nearly  constant value of 2.3$\AA$ (FWHM).   For each galaxy,
  we   degrade  the  MILES   models  (dashed   curve)  to   match  the
  corresponding wavelength dependent resolution (solid curves; see the
  text).~\label{RESOLUTION}}
\end{center}
\end{figure}

One main issue  for the estimate of velocity  dispersions from SDSS is
the  wavelength variation  of the  SDSS spectral  resolution.  For the
ETG's spectra of the SPIDER sample,  we found that the median value of
the resolution  varies from  $\sim 2.8\rm \AA$  (FWHM) ($\sigma_{inst}
\sim 90$~km/s)  in the  blue (4000$\rm \AA$)  up to $\sim  3.7\rm \AA$
(FWHM)   ($\sigma_{inst}   \sim  55$~km/s)   in   the  red   (8000$\rm
\AA$). Resolution  also varies significantly  among different spectra,
as seen in Fig.~\ref{RESOLUTION}, where we plot the $\sigma_{inst}$ as
a function  of wavelength  for the spectra  of five  randomly selected
ETGs. ~\citet{BER03a} have accounted  for the wavelength dependence of
the  SDSS spectral  resolution by  modeling  it with  a simple  linear
relation.
\begin{figure}
\begin{center}
\includegraphics[height=80mm]{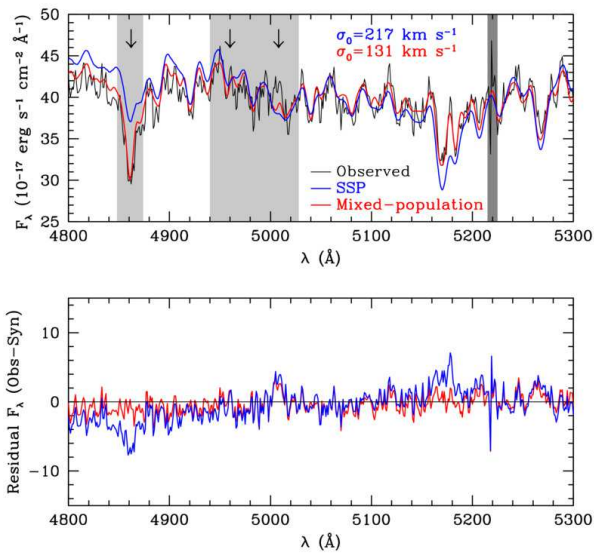}
\caption{ Example of the advantage in using mixed-population templates
  to derive  the galaxy central  velocity dispersion. The  upper panel
  shows an  example ETG  spectrum from SDSS  (J154615.45-001025.4) and
  the best-matching synthetic models obtained by running STARLIGHT (i)
  with a set  of 132 different SSP MILES models  (red color), and (ii)
  with  a  single SSP  model,  having an  age  of  12.6~Gyr and  solar
  metallicity (blue color). The  $\sigma_0$ values obtained in the two
  cases are  reported in  the upper panel.  The lower panel  shows the
  residuals  obtained  by  subtracting  the  models  to  the  observed
  spectrum. The gray bands in  the upper panel show the masked windows
  we  use   to  exclude  from  the  fitting   those  regions  possibly
  contaminated by emission lines (see  the arrows in the plot) as well
  as    corrupted     pixels    (e.g.    those     corresponding    to
  bad-columns).~\label{TEMPLATES}}
\end{center}
\end{figure} 
On the contrary, in the present  study, we do not perform any modeling
of $\sigma_{inst}$. For  each galaxy in the SPIDER  sample, we degrade
the   SSP  models  to   match  the   wavelength-dependent  resolution,
$\sigma_{inst}(\lambda)$,   of   the   corresponding  spectrum.    The
$\sigma_{inst}$ is measured  from the SDSS pipeline by  using a set of
arc lamps,  and provided  in one fits  extension of the  spectrum fits
file  (see~\citealt{EDR}).  The  MILES  models  are  degraded  by  the
transformation:
\begin{equation}
 M_s(\lambda)= \int M(x) G(x-\lambda) d\!x
\label{eq:conv_lambda}
\end{equation}
where  $M(\lambda)$  is  a  given  SSP model,  $M_s(\lambda)$  is  the
smoothed  model, and the  function $G(\lambda)$  is a  Gaussian kernel
whose  width  is  obtained  by  subtracting in  quadrature  the  MILES
resolution        to        the        de-redshifted        resolution
$\sigma_{inst}(\lambda/(1+z))$, where $z$  is the galaxy spectroscopic
redshift. The  integral is  performed by discrete  integration. Notice
that Eq.~\ref{eq:conv_lambda}  reduces to a simple  convolution in the
case  where $\sigma_{inst}$  is a  constant. For  each galaxy,  we run
STARLIGHT using the corresponding smoothed MILES models.  We use a set
of  SSP models covering  a wide  range of  age and  metallicity values
(with fixed solar [$\alpha$/Fe]=0  abundance ratio).  Age values range
from  $0.5$  to  $\sim  18$~Gyr,  while metallicity  values  of  $\log
Z/Z_{\odot}=-1.68,  -1.28,  -0.68,  -0.38,  0., 0.2$  are  considered,
resulting  in  a  total   of  132  SSP  models.   Fig.~\ref{TEMPLATES}
illustrates the  advantage of using  a mixed-population rather  than a
single  stellar  population template.   In  order  to exacerbates  the
difference between the two approaches, we selected the spectrum of one
ETG for which STARLIGHT measures a significant contribution from young
(age$<2$~Gyr) stellar populations.  The  Figure plots a portion of the
spectrum, along  with two best-fitting  models obtained by  either the
mixed-population  approach (in  red) or  by running  STARLIGHT  with a
single old stellar population template having an age of $12.6$~Gyr and
solar metallicity (in blue). Residuals are plotted for both cases.  It
is evident that the mixed-population model yields a better description
of  the   continuum  and  the   absorption  features  in   the  galaxy
spectrum. In particular, one may  notice that the $Mgb$ band ($\lambda
\sim  5170 \rm\AA$), which  is one  of the  main spectral  features to
measure the  $\sigma_0$, shows significantly smaller  residuals in the
case  of the  mixed-population  fit.  In  fact,  the $\sigma_0$  value
(reported  in   the  upper  panel   of  Fig.~\ref{TEMPLATES})  changes
dramatically from  one case  to the other.   Fig.~\ref{TEMPLATES} also
shows  the masked  regions  used to  avoid  either corrupted  spectral
regions  (e.g.,  bad  columns)  or regions  possibly  contaminated  by
nebular emission.   In particular, we show three  masked regions.  The
ones  at $\lambda  \sim$ 4850  and  4980 $\rm\AA$  avoid the  H$\beta$
(4861$\rm\AA$)   and   [OIII]   (4959   and  5007)   emission   lines,
respectively,  while the  one at  $\sim$5220 $\rm\AA$  excludes pixels
contaminated by a bad-column, as flagged in the SDSS spectrum.

\begin{figure}
\begin{center}
\includegraphics[height=80mm]{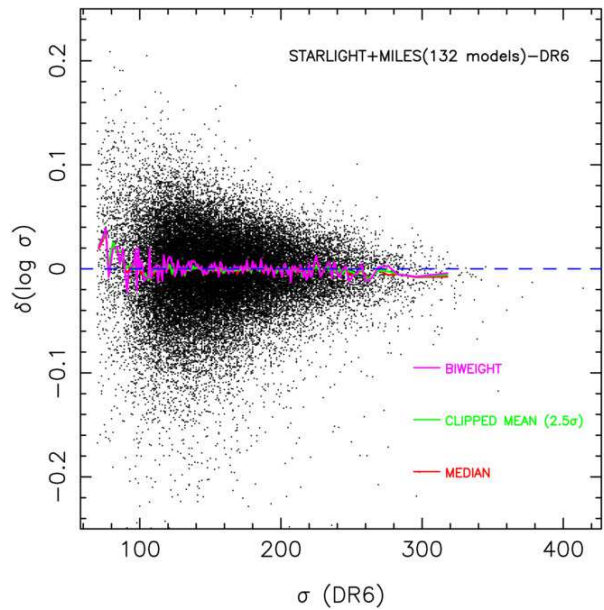}
\caption{Differences  in  \ls  \,  between  STARLIGHT and  SDSS  as  a
  function  of  the  SDSS  velocity  dispersion  value.   Curves  with
  different colors  are obtained by  binning the data with  respect to
  $\sigma_0$(DR6), and  taking for  each bin the  corresponding median
  value (red), the location  value estimated through the the bi-weight
  statistics  (magenta),  and   the  2.5$\sigma$  clipped  mean  value
  (green). Notice that the trend with $\sigma$ is essentially the same
  for all the three estimators.  The horizontal blue dashed line marks
  the value of  zero, corresponding to a null  difference between SDSS
  and STARLIGHT $\sigma_0$ values.~\label{VELDISP_CONF}}
\end{center}
\end{figure}

{  Fig.~\ref{VELDISP_CONF}   compares  velocity  dispersion  values
  obtained from the SDSS spectroscopic pipeline with those measured in
  this work using  STARLIGHT.  A good agreement is  found, with only a
  small  systematic  trend  at  the  low ($<  90$km/s)  and  high  ($>
  280$km/s)  ends  of  the   $\sigma_0$  range.   In  particular,  for
  $\sigma_0<90$km/s, the  STARLIGHT velocity dispersions  are slightly
  higher, by a few percent, with  respect to those of the SDSS.  It is
  important to  emphasize that  although STARLIGHT does  not normalize
  the spectrum by  the continuum, which is done  by the direct fitting
  method used in  the SDSS pipeline, the good  agreement found here is
  likely reflecting the excellent quality of flux calibration obtained
  in  DR6 (see  Figure  7 of~\citealt{DR6}).The  impact  of the  above
  systematic difference in $\sigma_0$ on the scaling relations of ETGs
  is investigated in paper II.}

\subsection{Uncertainties on the velocity dispersions}
\label{sec:err_veldisp}
\begin{figure}
\begin{center}
\includegraphics[height=80mm]{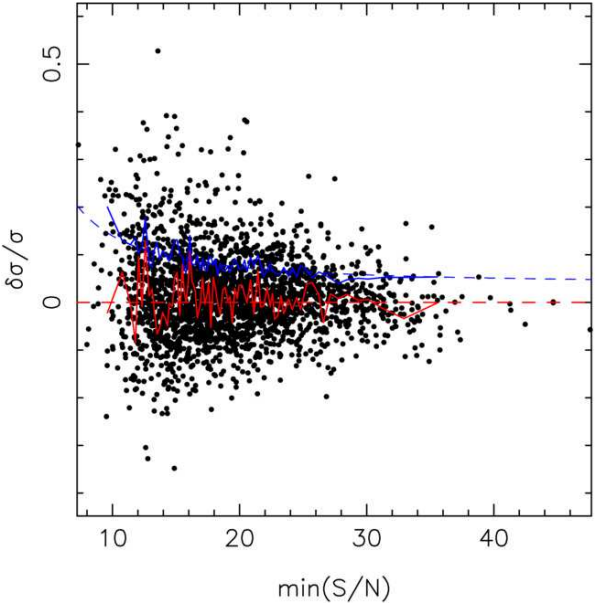}
\caption{Relative   differences    of   $\sigma_0$   values,   $\delta
  \sigma/\sigma$,  for   ETG's  spectra  with   repeated  measurements
  available  from  SDSS. Each  relative  difference  is computed  with
  respect  to the  value obtained  for the  spectrum with  highest S/N
  ratio. On  the x-axis,  the minimum S/N  ratio, $min(S/N)$,  of each
  pair of  spectra is  plotted. The red  solid curve shows  the median
  difference values  obtained by binning  the $\delta \sigma/\sigma$'s
  with respect to $min(S/N)$, with  each bin including the same number
  of 30 galaxies. The red dashed  curve marks the value of zero in the
  plot. The blue solid curve  shows rms of difference's values in each
  bin. The dashed blue line is  obtained by modeling the rms values by
  Eq.~\ref{eq:err_sigmas}.  ~\label{fig:diff_sigmas}}
\end{center}
\end{figure}

\begin{figure}
\begin{center}
\includegraphics[height=80mm]{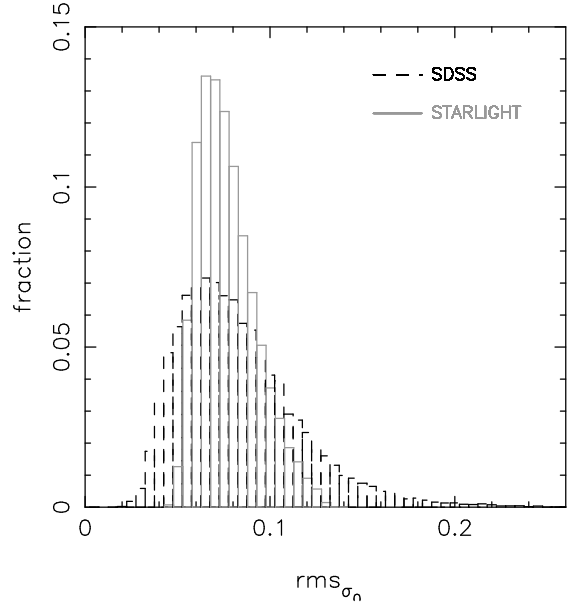}
\caption{Distribution  of  uncertainties  on SDSS  (dashed-black)  and
  STARLIGHT (gray)  $\sigma_0$ values. A small offset  of $+0.004$ has
  been  applied to the  gray histogram  to make  the plot  more clear.
  ~\label{fig:err_sigmas}}
\end{center}
\end{figure}

To estimate  the uncertainties on the STARLIGHT  $\sigma_0$ values, we
looked  for ETGs  in the  SPIDER sample  having spectra  with repeated
observations in SDSS.  Out of all the 39,993  galaxies, we found 2,313
cases with  a duplicate spectrum  available.  For all  these duplicate
spectra, we measure the $\sigma_0$ with STARLIGHT, with the same setup
and set of  model templates as for the primary  spectra. In each case,
we compute the  relative difference of $\sigma_0$ as  $\delta \sigma /
\sigma  =  (\sigma_{0,2}  -  \sigma_{0,1} )/\sigma_{0,1}$,  where  the
indices $1$  and $2$ refer  to the spectra  with higher and  lower S/N
ratios.  Fig.~\ref{fig:diff_sigmas} plots the $\delta \sigma / \sigma$
as a  function of the minimum  S/N ratio, $min(S/N)$, of  each pair of
duplicate spectra.  The S/N is computed  from the median  S/N ratio in
the  spectral region  of the  $\rm  H_{\rm \beta}$  feature, within  a
window of $100 \rm\AA$, centered at $\lambda=4860 \rm \AA$. We bin the
$\delta \sigma / \sigma$ values  with respect to $min(S/N)$, with each
bin including $30$  galaxies. For each bin, we  compute the median and
rms    values    of    $\delta    \sigma    /    \sigma$.    As    the
Fig.~\ref{fig:diff_sigmas}  shows, the median  values (red  color) are
fully  consistent   with  zero,   implying  that,  as   expected  (see
Sec.~\ref{sec:DATA}), the S/N ratio of  the spectra is large enough to
obtain  unbiased  velocity   dispersion  estimates.  The  rms  values,
$rms_{\sigma_0}$,  provide  an  estimate  of the  error  on  STARLIGHT
$\sigma_0$ values. The $rms_{\sigma_0}$ increases at low S/N ratio and
is  well described,  as  shown in  Fig.~\ref{fig:diff_sigmas}, by  the
following functional form:
\begin{equation}
 rms_{\sigma_0}= 0.0383 + 3.2 \times min(S/N)^{-1.5}.
\label{eq:err_sigmas}
\end{equation}
For  each galaxy  in the  SPIDER sample,  we assign  the error  on the
STARLIGHT  $\sigma_0$  value   using  the  above  equation,  replacing
$min(S/N)$ with  the median S/N  ratio in the $H_{\beta}$  region (see
above)  of  the  corresponding   galaxy  spectrum.  Notice  that  this
procedure assumes  that the  error on $\sigma_0$  depends only  on the
$S/N$ ratio, and is the same, for a given $S/N$, for both low and high
$\sigma_0$  galaxies.  In  fact,  we verified  that  considering  only
galaxies with $\sigma_0<130 \, km \, s^{-1}$, the dashed blue curve in
Fig.~\ref{fig:diff_sigmas}  does  not  change significantly,  shifting
upwards by  less than  $2 \%$. Fig.~\ref{fig:err_sigmas}  compares the
distribution  of uncertainties  of the  SDSS and  STARLIGHT $\sigma_0$
values.   The STARLIGHT  distribution is  significantly  narrower than
that of SDSS. This  is likely due to the fact that  we are using a one
parameter (i.e.  the $S/N$ ratio)  function to assign the  errors (see
Eq.~\ref{eq:err_sigmas}).    However,  the   peak  position   of  both
distributions  is very  similar. In  fact, the  median values  of SDSS
errors   and  STARLIGHT   $rms_{\sigma_0}$'s  are   fully  consistent,
amounting to $\sim 0.07 \rm dex ~(\sim 15\%$).

\section{Completeness}
\label{sec:compl}

\begin{figure}
\begin{center}
\includegraphics[width=70mm]{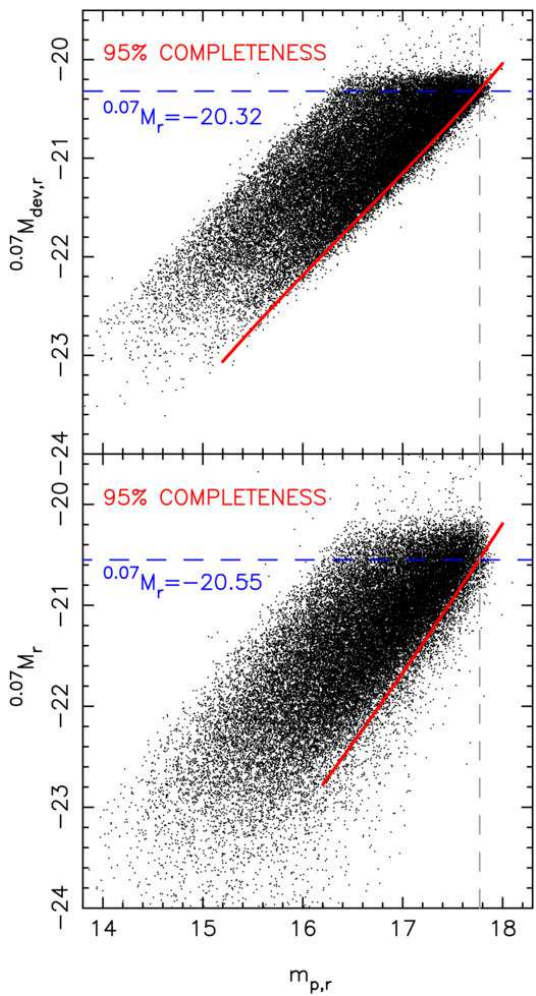}
\caption{Completeness limits  of the ETG sample for  the SDSS absolute
  model magnitudes (upper-panel) and 2DPHOT absolute Sersic magnitudes
  (lower-panel)  in r-band.   The vertical  dashed gray  line  in each
  panel marks  the Petrosian magnitude limit,  $m_{p,r}=17.77$, of the
  SDSS spectroscopy.  The red curve is obtained by a second polynomial
  fit of  the $95\%$ percentile values of  the $m_{p,r}$ distributions
  in different bins of total  magnitude (see the text). The horizontal
  dashed lines mark the points where the red curves cross the vertical
  line,   defining    the   $95\%$   completeness    limits   of   the
  sample. ~\label{MAG_LIMS}}
\end{center}
\end{figure} 

\subsection{SDSS vs. 2DPHOT completeness in $r$ band}
\label{sec:compl_r}
We first  analyze the completeness of  the SPIDER sample  in $r$ band.
The completeness  of SDSS data  is well characterized with  respect to
the Petrosian  magnitude, \mpetr, which  is one of the  main selection
criteria used to target objects  for SDSS spectroscopy. Here, we adopt
\mpetr$=17.77$   as  the   reference  value   for   the  spectroscopic
completeness~\citep{Strauss:2002}.   On   the  other  hand,  effective
parameters are defined in terms  of either the model magnitudes (SDSS)
or Sersic total  magnitudes (2DPHOT).  As shown in~\citet{Blanton:01},
the  difference  between Petrosian  and  model  magnitudes depends  on
galaxy    half-light    radii.      Moreover,    as    discussed    in
Sec.~\ref{sec:conf_SDSS_2DPHOT},    there    are   large    systematic
differences  between \photo  \, model  magnitudes, $m_{dev}$,  and the
2DPHOT Sersic total magnitudes, $m_T$.  This implies that the \mpet \,
limit maps  into a different  completeness magnitude value  when using
either SDSS  or 2DPHOT  parameters.  In the  following, we  denote the
absolute \mdev \, magnitude as  \mrdev, and total 2DPHOT magnitudes as
\mr, characterizing the completeness of the SPIDER sample with respect
to both  \mrdev \,  and \mr.  We  also use  the symbol $M_r$  to refer
indistinctly   to   either  one   or   the   other  total   magnitude.
Fig.~\ref{MAG_LIMS}  plots SDSS  and 2DPHOT  absolute magnitudes  as a
function of \mpetr.   At a given \mpetr, the scatter  seen in the plot
reflects all the different factors  that enter the definition of total
absolute magnitudes, i.e.  the k- and galactic extinction corrections,
the redshift range of the  sample, as well as the intrinsic difference
between $M_r$ and  \mpet \, (see above).  Assuming  the SDSS sample to
be  complete   down  to  \mpetr$=17.77$,   we  define  here   $95  \%$
completeness limits for  \mrdev \, and \mr, by  adopting the geometric
approach illustrated  in the  figure.  The method  is similar  to that
described by ~\citet{GMA99}, where  the completeness limit of a galaxy
sample is defined as that  magnitude where galaxies begin to be missed
in it because of the  surface brightness detection limit. In practice,
we select a range in absolute magnitude where all galaxies have \mpetr
\,  smaller   than  the  SDSS  spectroscopic   completeness  limit  of
$17.77$. We  consider the  ranges of $-20.7$  to $-21.7$ and  $-21$ to
$-22$ for \mrdev \, and  \mr, respectively.  Then, in these ranges, we
bin the  distribution of \mpetr \,  values with respect  to $M_r$, and
derive the  $95\%$ percentile  of the \mpetr  \, distribution  in each
bin.  The  binned points  are fitted with  a second  order polynomial,
shown by  the red curves  in Fig.~\ref{MAG_LIMS}.  The $M_r$  \, value
where the polynomial intersects the  vertical line of \mpetr$ = 17.77$
defines the  point below  which at least  $95\%$ of the  galaxies, for
whatever value of $M_r$, are included  in the sample.  We refer to the
value of  $M_r$ at the  intersection point as the  $95\%$ completeness
limit  of the  sample. Notice  that the  2DPHOT completeness  limit is
brighter  than that of  the model  magnitudes.  The  difference, $\sim
0.2$mag, matches the  average difference between \mrdev \,  and \mr \,
measured  for  the  faintest   galaxies  in  the  SPIDER  sample  (see
Sec.~\ref{sec:conf_SDSS_2DPHOT}).

\begin{figure*}[!ht]
\begin{center}
\includegraphics[width=140mm]{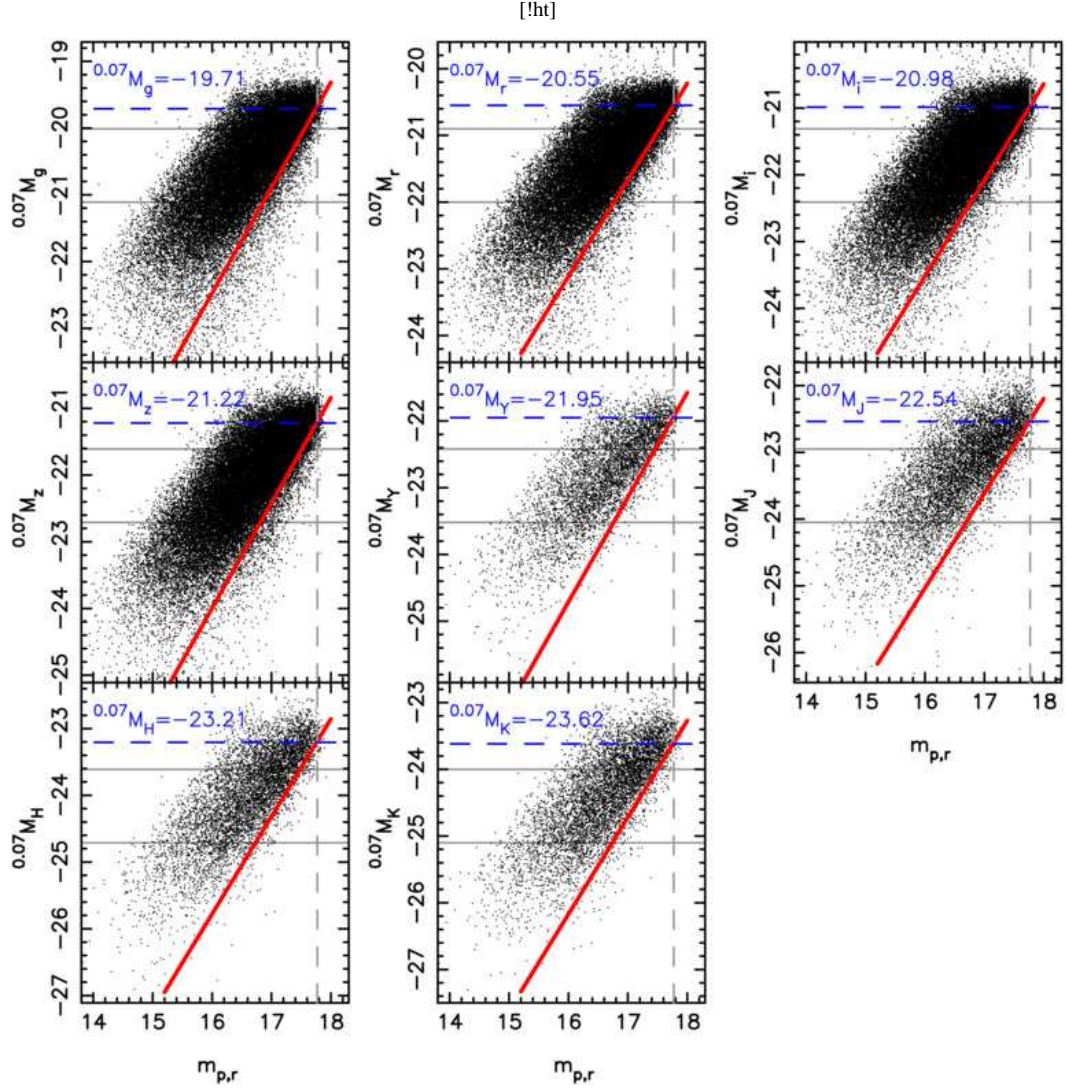}
\caption{Completeness limits  in the  $grizYJHK$ bands. Each  panel is
  the  same as  the  lower  panel in  Fig.~\ref{MAG_LIMS},  but for  a
  different waveband,  $X$. The  $X$ runs from  $g$ to $K$  going from
  left to right, and top to  bottom in the figure. The upper-mid panel
  is  the same  as  the  lower panel  of  Fig.~\ref{MAG_LIMS}, and  is
  repeated here for a better  comparison among all the wavebands.  The
  $95\%$  completeness  limits (see  the  text)  are  reported in  the
  upper--left  corners of the  panels. The  two solid  gray horizontal
  lines  indicate  the  absolute  magnitude  range  of  \mx$_{_1}$  to
  \mx$_{_2}$,   where  the   sample  is   complete  with   respect  to
  \mpetr,~\label{MAG_LIMS_ALL}}
\end{center}
\end{figure*} 

\subsection{Completeness from $g$ through $K$}
We   apply  the  geometric   approach  of   Sec.~\ref{sec:compl_r}  to
characterize  the  completeness limit  of  the  SPIDER  sample in  all
wavebands. Since  our main  goal is that  of selecting  volume limited
samples for analyzing the FP  relation in each waveband (papers II and
III),  we  consider here  only  2DPHOT  Sersic  magnitudes, which  are
linearly  related to  \lre \,  and \mie,  and hence  are  the relevant
quantities to characterize the completeness of the sample in the space
of  effective  parameters.   Fig.~\ref{MAG_LIMS_ALL} plots  the  total
Sersic magnitudes, \mx,  \, as a function of \mpetr,  where $X$ is one
of the available wavebands (i.e.   $X=grizYJHK$). For a given band, we
define an absolute magnitude  range of \mx$_{_1}$ to \mx$_{_2}$, where
the  sample  is   complete  with  respect  to  \mpetr,   and  bin  the
corresponding  distribution of \mpetr  \, values  with respect  to the
\mx. The $95\%$ completeness magnitude in the band $X$ is then defined
as described in Sec.~\ref{sec:compl_r}.   For each band, the values of
\mx$_{_1}$ and \mx$_{_2}$ are  obtained by trasforming those in r-band
through   the  median  values   of  the   ETG's  color   indices  (see
Tab.~\ref{tab:meancol}).  For each band, the $95\%$ completeness limit
is  reported  in Tab.~\ref{tab:compl},  together  with  the number  of
SPIDER ETGs brighter than that limit.

\begin{table}
%\centering
\small
\begin{minipage}{50mm}
 \caption{Completeness limits in $grizYJHK$.
Column 1: waveband. Column 2: $95\%$ completeness magnitude. Column 3:
number of galaxies brighter than the completeness limit.}
  \begin{tabular}{c|c|c}
   \hline
waveband       & $^{0.07}M_{X}$ limit  & $N$  \\
\hline
$g$ & $ -19.71$ & $35989$ \\
$r$ & $ -20.55$ & $36205$ \\
$i$ & $ -20.98$ & $35711$ \\
$z$ & $ -21.22$ & $36310$ \\
$Y$ & $ -21.95$ & $4652$ \\
$J$ & $ -22.54$ & $6432$ \\
$H$ & $ -23.21$ & $5823$ \\
$K$ & $ -23.62$ & $5690$ \\
   \hline
  \end{tabular}
\label{tab:compl}
\end{minipage}
\end{table}

\section{Summary}
\label{sec:summary}

The data  presented in this  paper represents the most  extensive ever
obtained for early-type galaxies. Optical data from SDSS were entirely
reprocessed; images with 2DPHOT (La  Barbera et al.  2008) and spectra
with STARLIGHT, as described above. This allows us to perform a proper
comparison of different pipelines. We have created a database to store
all of  the photometric and spectroscopic parameters  measured as part
of our reprocessing. The SPIDER database (SdB) currently contains only
the  photometric  (SDSS  and  UKIDSS) and  spectroscopic  (SDSS)  data
described  here, but  we expect  to incorporate  additional  data from
other  wavelength  regimes. The  SdB  architecture  is extensible  and
designed  to  support  this  growing  process.   {  We  are  currently
  developing  an intuitive  Graphic  User Interface  (GUI) which  will
  allow the user  to easily retrieve all information  available in the
  SdB issuing SQL  queries. Until the GUI is ready  for general use we
  are      making     the      data     available      through     the
  link\\ http://www.lac.inpe.br/bravo/arquivos/SPIDER\_data\_paperI.ascii.}\footnote{A
  mirror           is            also           available           at
  http://www.na.astro.it/~labarber/SPIDER/.}

The  data  presented here  will  be  used  in the  forthcoming  papers
analyzing the FP relations (papers  II and III).  Given the importance
of  obtaining  meaningful measurements  of  the structural  parameters
entering  the  FP, this  contribution  examines  how consistent  these
parameters are and their errors. Here we summarize the main properties
of the sample defined in  this paper and the characteristic parameters
of each ETG:

1 - When  matching optical$+$NIR data we end-up  with 5,080 ETGs which
can then be used to study the global properties of elliptical galaxies
in the nearby Universe (z$<$0.1).

2  - For  each  of these  ETGs  we have  measured  aperture and  total
magnitudes,  k-corrected and  dereddened for  galactic  extinction. We
also  quantify for  each  galaxy the  following essential  parameters:
effective   (half-light)  radius,  $r_{\rm   e}$,  the   mean  surface
brightness  within this  radius, $<  \!\mu\!>_{\rm e,  }$,  the Sersic
index  (shape  parameter) $\rm  n$,  the  axis  ratio $b/a$,  and  the
position angle of the major  axis, $PA$. Besides, the galaxy isophotal
shape is characterized by the a$_{4}$ parameter.

3  - Uncertainties  in all  of  the parameters  previously listed  are
estimated  and  presented  as  a  function of  the  logarithm  of  the
signal-to-noise  per   pixel.  Median   errors  in  $r_{\rm   e}$,  $<
\!\mu\!>_{\rm   e,   }$,  and   $\rm   n$   are   $\sim  0.1$,   $\sim
0.5$~$mag/arcsec^2$, and $\sim 0.1$, respectively, in the optical, and
$\sim 0.15$, $\sim 0.6$~$mag/arcsec^2$, and $\sim 0.12$ in the NIR.

4 - We  do not find any systematic steepening of  the CM relation when
considering a  waveband baseline from  $g-r$ to $g-K$, as  expected if
the {\it  total} CM relation  is solely driven by  a mass--metallicity
relation in  ETGs. According to  Scodeggio (2001), this result  can be
interpreted as follows: ETGs have negative color gradients, with color
indices becoming bluer from the  galaxy center to its outskirts. Thus,
when we adopt colors in a  fixed aperture, we measure a color inside a
smaller region (with respect to  \re) for the brightest (hence larger)
galaxies  than for  faintest  galaxies  in the  sample,  leading to  a
misleading steepening of the CM relation

5  - The study  presented here  of the  structural parameters  of ETGs
reveals  the following properties:  log $\rm  r_{\rm e}$  decreases by
35\%  from the  optical  to  the NIR,  reflecting  the internal  color
gradients  in  these  systems;   the  axis  ratios,  b/a,  have  their
distribution  essentially constant from  $g$ through  K, with  a media
value of $\sim$0.7 and width  of $\sim$0.2; The Sersic index is always
larger than  1, since  we selected only  bulge dominated  galaxies and
spans a  domain from $\sim$2 to $\sim$10,  with a median of  6 for all
wavebands; no systematic trend of a$_{4}$ was found from $g$ to $i$.

6 -  We present measurements  of central velocity  dispersion obtained
using STARLIGHT.  Extensive comparison with the  estimates provided by
the  SDSS pipeline,  shows that  our estimates  are unbiased  over the
whole $\sigma_{\circ}$ range, with a median error of $\sim$15\%.

7 - Comparison of two independent pipelines was done (SDSS and 2DPHOT)
and reveals significant differences  in magnitude and effective radius
with  respect to  the Sersic  index  and absolute  model magnitude  in
r-band. The  impact of  such differences on  the scaling  relations of
ETGs will be addressed in paper II.

\section*{Acknowledgments}
We thank  the staffs in  charge of the  clusters at the  INPE-LAC (Sao
Jos\'e dos  Campos, Brazil), H.C.   Velho, and the staffs  at INAF-OAC
(Naples,  Italy), Dr.  A.Grado  and F.I.Getman,  for keep  running the
systems smoothly. We also thank M. Capaccioli for the support provided
to  this  project.  We  thank  R.   Gal  for several  suggestions  and
comments throughout this project.   We thank M.Bernardi for helping us
to retrieve  velocity dispersions  from SDSS. {  We also  thank the
  anonymous  referee for  the helpful  comments and  suggestions}.  We
have used data  from the 4th data release of  the UKIDSS survey, which
is  described  in detail  in  \citet{War07}.   The  UKIDSS project  is
defined  in~\citet{Law07}.  UKIDSS  uses the  UKIRT Wide  Field Camera
(WFCAM; Casali et  al, 2007).  The photometric system  is described in
Hewett et  al (2006), and the  calibration is described  in Hodgkin et
al.  (2009). The pipeline processing and science archive are described
in Irwin et  al (2009, in prep) and Hambly et  al (2008).  Funding for
the  SDSS  and SDSS-II  has  been provided  by  the  Alfred P.   Sloan
Foundation,  the  Participating  Institutions,  the  National  Science
Foundation, the  U.S.  Department of Energy,  the National Aeronautics
and Space Administration, the  Japanese Monbukagakusho, the Max Planck
Society, and  the Higher Education  Funding Council for  England.  The
SDSS Web  Site is  http://www.sdss.org/.  The SDSS  is managed  by the
Astrophysical Research Consortium  for the Participating Institutions.
The  Participating Institutions  are  the American  Museum of  Natural
History,  Astrophysical   Institute  Potsdam,  University   of  Basel,
University of  Cambridge, Case Western  Reserve University, University
of Chicago,  Drexel University,  Fermilab, the Institute  for Advanced
Study, the  Japan Participation  Group, Johns Hopkins  University, the
Joint  Institute for  Nuclear  Astrophysics, the  Kavli Institute  for
Particle Astrophysics  and Cosmology, the Korean  Scientist Group, the
Chinese Academy of Sciences  (LAMOST), Los Alamos National Laboratory,
the     Max-Planck-Institute     for     Astronomy     (MPIA),     the
Max-Planck-Institute   for  Astrophysics   (MPA),  New   Mexico  State
University,   Ohio  State   University,   University  of   Pittsburgh,
University  of  Portsmouth, Princeton  University,  the United  States
Naval Observatory, and the University of Washington.

%Bujarrabal  et al. (1981),  Efstathiou \&
%Fall (1984), Whitmore (1984), and Okamura et al. (1984)

\bsp

\label{lastpage}

\end{document}